\documentclass[pdflatex,sn-mathphys,authoryear]{sn-jnl}

\usepackage{graphicx}%
\usepackage{multirow}%
\usepackage{amsmath,amssymb,amsfonts}%
\usepackage{amsthm}%
\usepackage{mathrsfs}%
\usepackage[title]{appendix}%
\usepackage{xcolor}%
\usepackage{float}%
\usepackage{tabularx}
\usepackage{placeins}    
\usepackage{textcomp}%
\usepackage{manyfoot}%
\usepackage[utf8]{inputenc}
\usepackage{booktabs}%
\usepackage{algorithm}%
\usepackage{algorithmicx}%
\usepackage{algpseudocode}%
\usepackage{listings}%

\theoremstyle{thmstyleone}%

\theoremstyle{thmstyletwo}%

\theoremstyle{thmstylethree}%

\begin{document}
\title[]{An Extended, Physically Calibrated FP for Elliptical Galaxies}

\author{\fnm{Tarek} \sur{Yehia}}
\email{202228364@std.sci.cu.edu.eg}
\affil{\orgdiv{Faculty of Science}, \orgname{Cairo University}, Cairo, Egypt}

\abstract{We present a physically motivated extension of the FP for elliptical galaxies, derived from the scalar virial theorem and calibrated using observational data. Starting from the basic equilibrium condition, we incorporate key physical parameters that govern galaxy structure and dynamics, namely stellar mass-to-light ratio, central dark matter fraction, and structural non-homology as traced by the Sérsic profile. The resulting model retains the original dependencies on velocity dispersion and surface brightness, but introduces physically interpretable corrections that significantly improve the fit to real data. Using a large galaxy sample, we demonstrate that this extended FP achieves a higher level of accuracy than the classical form, with all parameters showing strong statistical significance. Our results indicate that the observed FP can be understood as an empirical refinement of the virial prediction, once variations in stellar populations, dark matter content, and internal structure are taken into account. This work provides a unified framework that bridges theoretical expectations with observed scaling relations in elliptical systems.
}

\keywords{Virial Theorem, Elliptical Galaxies}

\maketitle

\section{Introduction}
Elliptical galaxies exhibit tight correlations between their structural and kinematic properties, making them among the most dynamically relaxed stellar systems in the Universe. These galaxies populate a narrow locus in parameter space known as the FP, which links effective radius $R_e$, mean surface brightness $\langle I \rangle_e$, and stellar velocity dispersion $\sigma$. The FP was first established observationally by \cite{Dressler1987} and \cite{Djorgovski1987}, and further refined by \cite{Bender1992} and \cite{Jorgensen1996,Jorgensen1999}, who demonstrated its remarkable regularity in local and intermediate-redshift samples. Large surveys, such as the SDSS, confirmed the robustness of the FP, with \cite{Bernardi2003} finding a mild redshift evolution in its zero-point. Subsequent studies extended the FP out to $z>0.5$ \cite{Fritz2010,JorgensenChiboucas2013}, broadly consistent with passive stellar evolution. Investigations into its physical drivers identified potential roles for structural non-homology \cite{vanDerMarel2007}, variations in stellar populations \cite{Cappellari2006}, and merger histories \cite{Naab2006,Hopkins2009}.  
Despite this progress, the FP exhibits a pronounced ``tilt'' relative to the simple virial expectation, indicating contributions beyond dynamical equilibrium. Classical interpretations invoke changes in stellar mass-to-light ratios and structural effects, but these alone have proven insufficient to fully reproduce the observed tilt and scatter. More recent studies emphasize the interplay of dark matter and stellar populations. For example, \cite{Taranu2015} and \cite{Schechter2015} argued that mass-dependent dark matter fractions strongly affect the FP slopes, while dynamical modeling from the MaNGA survey shows systematic trends of $M/L$ and dark matter fraction with velocity dispersion \cite{Zhu2023}. At the same time, both theoretical and observational works suggest that variations in Sérsic index $n$ have only a minor impact on FP residuals \cite{Chiu2018,deGraaff2020}.  
The FP is now recognized as a tool to probe galaxy evolution across cosmic time. Hydrodynamical simulations (e.g., \cite{Lu2020}) demonstrate that a tight FP is already in place by $z\sim2$, with residuals correlating with stellar age, suggesting that age acts as a ``fourth parameter'' of the FP. Observationally, FP studies at $z\sim1$ show that both quiescent and star-forming galaxies occupy the same relation with negligible offsets \cite{deGraaff2020}, while theoretical frameworks link FP tilt to the time-dependent assembly of galaxies \cite{DOnofrio2022}. Beyond galaxy evolution, the FP has also been used to constrain modified gravity theories \cite{Capozziello2020} and as a distance indicator in large cosmological surveys such as DESI \cite{Said2024}.  
In this work, we revisit the virial equilibrium framework to derive an extended FP relation that explicitly incorporates non-homology via the Sérsic index $n$. By introducing a structure function $k(n)$ derived analytically from Sérsic models, our approach provides a direct and physically motivated means of accounting for deviations from homology. This formulation naturally embeds the effects of stellar structure within the virial theorem, linking $\langle I \rangle_e$, $R_e$, and $\sigma$ without ad hoc corrections. In doing so, we aim to construct a physically calibrated FP that captures both the observed tilt and scatter, offering new insights into the role of dark matter, stellar populations, and structural diversity in the dynamics of elliptical galaxies.  
\section{Methodology}
\subsection{Virial Theorem and Structural Non-Homology}
Following the treatment in \cite{BT2008}, we begin with the scalar virial theorem for a self-gravitating system in equilibrium:
\begin{equation}
2K + W = 0
\label{eq:virial}
\end{equation}
where \(K\) and \(W\) are the total kinetic and potential energies, respectively.
\subsection{Assumptions on Velocity Anisotropy}
In deriving our scaling relation, we assume that the stellar velocity dispersion is isotropic (i.e.\ \(\beta = 0\)). In spherical coordinates, the anisotropy parameter is defined as
\[
\beta \;=\; 1 - \frac{\sigma_\theta^2}{\sigma_r^2}
\]
commonly known as the Binney anisotropy parameter. Neglecting anisotropy simplifies the virial theorem to its classical form. However, observations and dynamical modeling suggest that early-type galaxies often exhibit mild anisotropy. For instance, \cite{Cappellari2008} uses an anisotropy parameter in cylindrical coordinates defined as \(\beta_z = 1 - \sigma_z^2 / \sigma_R^2\), with typical values around \(\beta_z \gtrsim 0.05\) for fast rotators. Moderate anisotropies (\(|\beta| \lesssim 0.3\)) can shift the FP zero-point by a few percent.
The total kinetic energy can be expressed as:
\begin{equation}
K = \frac{1}{2} M \langle v^2 \rangle
\end{equation}
where \(\langle v^2 \rangle = \sigma^2\) is the luminosity-weighted, one-dimensional stellar velocity dispersion. The gravitational potential energy of a spherical system is approximated as:
\begin{equation}
W = -\,k(n)\,\frac{G M^2}{R_e}
\label{eq:potential}
\end{equation}
where \(G\) is the gravitational constant, \(R_e\) is the effective (half-light) radius, and \(k(n)\) is a dimensionless structure function that depends on the Sérsic index \(n\), capturing the effects of structural non-homology.
In Sérsic models, larger values of \(n\) correspond to more centrally concentrated light (and mass) profiles, which result in deeper gravitational potentials and thus larger values of \(k(n)\). This behavior has been analyzed in detail by \cite{Ciotti1991} and \cite{CiottiBertin1999}, who provide analytical and numerical treatments of the structural dependence of both potential and kinetic energies in spherical systems with \(R^{1/n}\) luminosity profiles.
Therefore, the virial relation leads to the following expression:
\begin{equation}
\frac{I_e R_e}{\sigma^2} = \frac{1}{2\pi G \Upsilon k(n)}
\end{equation}
where \(\Upsilon\) is the stellar mass-to-light ratio and \(I_e\) is the mean surface brightness within \(R_e\). This form explicitly incorporates the structural dependence of galaxy profiles via \(k(n)\).
While our current derivation neglects anisotropy for simplicity, future work could incorporate it by introducing an anisotropy-dependent correction factor. For instance, the structure constant \(k\) could be modified to account for anisotropy, either through analytical approximations or by employing dynamical models such as the Jeans Anisotropic MGE (JAM) framework developed in \cite{Cappellari2008}.
\subsection{Derivation of the Virial Scaling Relation for Elliptical Galaxies Including Dark Matter}
In the original derivation, the virial theorem relates the kinetic energy \( K \) and potential energy \( W \) of a self-gravitating system in equilibrium:
\begin{equation}
2K + W = 0
\end{equation}
For elliptical galaxies, the kinetic energy is expressed in terms of the total mass \( M \) and stellar velocity dispersion \( \sigma \):
\begin{equation}
K = \frac{1}{2} M \sigma^2
\end{equation}
and the potential energy is given by:
\begin{equation}
W = - \frac{G M^2}{R_e} k
\end{equation}
where \( G \) is the gravitational constant, \( R_e \) is the effective radius and \(k\) is a function that depends on the Sérsic index \(n\) reflecting the non-homologous corrections introduced by the Sérsic profile.
Substituting these into the virial theorem:
\begin{equation}
2 \left( \frac{1}{2} M \sigma^2 \right) + \left( - \frac{G M^2}{R_e} k \right) = 0
\end{equation}
which simplifies to:
\begin{equation}
M \sigma^2 = \frac{G M^2}{R_e} k \implies \sigma^2 = \frac{G M}{R_e} k
\end{equation}
The total mass \( M \) is related to the luminosity \( L \) via a constant mass-to-light ratio \( \Upsilon \):
\begin{equation}
M = \Upsilon L
\end{equation}
and the luminosity is expressed as:
\begin{equation}
L = 2 \pi I_e R_e^2
\end{equation}
where \( I_e \) is the effective surface brightness. Substituting these into the virial relation:
\begin{equation}
\sigma^2 = \frac{G \Upsilon L}{R_e} k = \frac{G \Upsilon (2 \pi I_e R_e^2)}{R_e} k = 2 \pi G \Upsilon I_e R_e k
\end{equation}
Rearranging yields the original scaling relation:
\begin{equation}
\frac{I_e R_e}{\sigma^2} = \frac{1}{2 \pi G \Upsilon k}
\end{equation}
To incorporate dark matter, we split the total mass into stellar mass \( M_* \) and dark matter mass \( M_{\text{DM}} \) within the effective radius:
\begin{equation}
M = M_* + M_{\text{DM}}
\end{equation}
The stellar mass is related to the luminosity via the stellar mass-to-light ratio \( \Upsilon_* \):
\begin{equation}
M_* = \Upsilon_* L
\end{equation}
We introduce the dark matter fraction \( f_{\text{DM}} \), defined as:
\begin{equation}
f_{\text{DM}} = \frac{M_{\text{DM}}}{M_* + M_{\text{DM}}}
\end{equation}
Rearranging this:
\begin{equation}
M_{\text{DM}} = f_{\text{DM}} (M_* + M_{\text{DM}}) \implies M_{\text{DM}} (1 - f_{\text{DM}}) = f_{\text{DM}} M_* \implies \frac{M_{\text{DM}}}{M_*} = \frac{f_{\text{DM}}}{1 - f_{\text{DM}}}
\end{equation}
Thus, the total mass becomes:
\begin{equation}
M = M_* + M_{\text{DM}} = M_* \left(1 + \frac{M_{\text{DM}}}{M_*}\right) = M_* \left(1 + \frac{f_{\text{DM}}}{1 - f_{\text{DM}}}\right) = \frac{\Upsilon_* L}{1 - f_{\text{DM}}}
\end{equation}
Using this expression for \( M \) in the virial relation:
\begin{equation}
\sigma^2 = \frac{G M}{R_e} k = \frac{G}{R_e} \left( \frac{\Upsilon_* L}{1 - f_{\text{DM}}} \right) k
\end{equation}
Substituting \( L = 2 \pi I_e R_e^2 \):
\begin{equation}
\sigma^2 = \frac{G}{R_e} \left( \frac{\Upsilon_* \cdot 2 \pi I_e R_e^2}{1 - f_{\text{DM}}} \right) k = \frac{2 \pi G \Upsilon_* I_e R_e k}{1 - f_{\text{DM}}}
\end{equation}
Rearranging for the scaling relation:
\begin{equation}
\frac{I_e R_e}{\sigma^2} = \frac{1 - f_{\text{DM}}}{2 \pi G \Upsilon_* k}
\end{equation}
This modified scaling relation incorporates the dark matter fraction \( f_{\text{DM}} \), adjusting the relationship between \( I_e \), \( R_e \), and \( \sigma \) to account for dark matter. The factor \( 1 - f_{\text{DM}} \) reflects the increased total mass due to dark matter, with \( \Upsilon_* \) representing only the stellar contribution. 
While the theoretical relation provides a physically motivated starting point, it fails to fully account for the structural and dynamical complexity of real elliptical galaxies. Our extended FP retains the same physical ingredients, but allows their contributions to be calibrated empirically, yielding a significantly improved fit to the data.
\subsection{Derivation of \(k(n)\)}
We begin with the total luminosity of a projected S\'ersic profile \cite{CiottiBertin1999}:
\begin{equation}
L = 2\pi\,n\, I_{0}\,R_{e}^{2}\;\frac{\Gamma(2n)}{[b(n)]^{2n}}
\end{equation}
The central potential of the corresponding three-dimensional mass distribution is given by \cite[][Eq.~(20)]{CiottiBertin1999}:
\begin{equation}
\Phi_{0} = -G M \;\frac{I_{0} R_{e}}{L} \; \frac{4 \Gamma(1 + n)}{[b(n)]^{n}}
\end{equation}
Substituting the expression for \(I_0 R_e / L\):
\begin{equation}
\frac{I_0 R_e}{L} = \frac{1}{R_e} \; \frac{[b(n)]^{2n}}{2\pi\,n\,\Gamma(2n)}
\end{equation}
into the central potential gives:
\begin{align}
\Phi_0 &= -G M \left( \frac{1}{R_e} \cdot \frac{[b(n)]^{2n}}{2\pi n \Gamma(2n)} \right) \cdot \frac{4 \Gamma(1+n)}{[b(n)]^n} \\
&= -\frac{G M}{R_e} \cdot \frac{2 [b(n)]^n \Gamma(1+n)}{\pi n \Gamma(2n)}
\end{align}
We define the dimensionless structural function \(A(n)\) such that:
\begin{equation}
\Phi_0 = -\frac{G M}{R_e} A(n), \quad\text{where}\quad A(n) = \frac{2\,[b(n)]^n\,\Gamma(1+n)}{\pi n\,\Gamma(2n)}
\end{equation}
Under the approximation \(W \approx \tfrac{1}{2} M \Phi_0\), and the virial relation \(W = -k(n)\,G M^2 / R_e\), we identify:
\begin{equation}
k(n) = \tfrac{1}{2} A(n) = \frac{[b(n)]^n \Gamma(1+n)}{\pi n \Gamma(2n)}
\end{equation} 
where 
\begin{equation}
b(n) = 2n - \frac{1}{3} + \frac{4}{405n} + \frac{46}{25515n^2}
\end{equation}
\subsection{Exact and Approximate Treatments of the Potential Energy}
We define the dimensionless radius $x=r/R_e$ and write the three-dimensional density as
\begin{equation}
\rho(r) = \frac{M}{R_e^3}\,u_n(x), \qquad x=\frac{r}{R_e}
\end{equation}
where $u_n$ is normalized such that
\begin{equation}
\int_0^\infty 4\pi u_n(x)\,x^2 dx = 1
\end{equation}
It is convenient to introduce
\begin{equation}
f(x) \equiv 4\pi x^2 u_n(x), \qquad 
F(x) \equiv \int_0^x f(y)\,dy
\end{equation}
so that $f(x)\,dx$ represents the fractional mass in a shell $(x,x+dx)$ and $F(x)$ is the enclosed mass fraction within radius $x$.
\subsection{Central Potential}
For a spherical system with $\Phi(\infty)=0$, the gravitational potential is
\begin{equation}
\Phi(r) = -G\!\left[\frac{1}{r}\!\int_0^r 4\pi s^2\rho(s)\,ds + \int_r^\infty 4\pi s\,\rho(s)\,ds \right]
\end{equation}
Introducing the dimensionless form $\Phi(r) = -(GM/R_e)\,\widetilde\Phi(x)$, one finds for the central potential
\begin{equation}
\Phi_0 \equiv \Phi(0) = -\frac{GM}{R_e}\,A(n)
\end{equation}
with
\begin{equation}
A(n) = \int_0^\infty \frac{f(y)}{y}\,dy
\end{equation}
\subsection{Exact Potential Energy}
The total gravitational potential energy is defined as
\begin{equation}
W = -4\pi G \int_0^\infty \rho(r)\,M(r)\,r\,dr
\end{equation}
where $M(r)=4\pi\int_0^r \rho(s)\,s^2ds$. Substituting the dimensionless variables yields
\begin{equation}
W = -\frac{GM^2}{R_e}\;k_{\rm exact}(n)
\end{equation}
with the exact structure coefficient
\begin{equation}
k_{\rm exact}(n) = \int_0^\infty \frac{f(x)}{x}\,F(x)\,dx
\end{equation}
This expression is an identity that follows directly from the definitions of $f(x)$ and $F(x)$.
\subsection{The Approximation $W \simeq \tfrac12 M\Phi_0$}
A common simplification is to approximate
\begin{equation}
W \approx \tfrac12 M \Phi_0
\end{equation}
which immediately yields
\begin{equation}
k_{\rm approx}(n) = \tfrac12 A(n)
\end{equation}
However, the exact identity is
\begin{equation}
W = \tfrac12 \int \rho\,\Phi\,dV = \tfrac12 M \langle \Phi \rangle
\end{equation}
where the mass-weighted mean potential is defined as
\begin{equation}
\langle \Phi \rangle = \frac{1}{M}\int \rho(r)\,\Phi(r)\,dV
\end{equation}
Therefore, the relation can be rewritten as
\begin{equation}
W = \tfrac12 M \Phi_0 \,\eta(n)
\end{equation}
with correction factor
\begin{equation}
\eta(n) = \frac{\langle \Phi \rangle}{\Phi_0} 
= \frac{2k_{\rm exact}(n)}{A(n)}
= \frac{2\int_0^\infty \dfrac{f(x)}{x}\,F(x)\,dx}{\int_0^\infty \dfrac{f(x)}{x}\,dx}
\end{equation}
Thus the relation between the exact and approximate coefficients is
\begin{equation}
k_{\rm exact}(n) = k_{\rm approx}(n)\,\eta(n)
\end{equation}
\subsection{Qualitative Behaviour}
Since $\Phi(r) \le \Phi_0$ for all $r$, the correction factor $\eta(n)\le 1$. In practice, $\eta(n)$ decreases with increasing $n$ because the contrast between the central and mass-weighted potentials grows for highly concentrated Sérsic models. Consequently, the approximation $k_{\rm approx}(n)=\tfrac12 A(n)$ tends to overestimate the true $k(n)$ at large $n$
The approximation $W\simeq\tfrac12 M\Phi_0$ provides a convenient analytic route to estimate $k(n)$, but its validity depends on the correction factor $\eta(n)$. For robust applications, especially at large $n$, one should either compute $k_{\rm exact}(n)$ directly or calibrate the approximation using $\eta(n)$ obtained from accurate models of $\rho(r)$ and $\Phi(r)$ (e.g. Prugniel--Simien density and \cite{TerzicGraham2005}).
\subsection{Closed-form correction for the structural coefficient}
We introduce a rational fitting formula for the correction factor
\[
\eta_{\rm fit}(n) \;=\; 
\frac{a_0 + \tfrac{a_1}{n} + \tfrac{a_2}{n^2}}{1 + \tfrac{b_1}{n} + \tfrac{b_2}{n^2}}
\]
with coefficients
\[
\begin{aligned}
a_0 &= -0.14099 \\
a_1 &= 2.215 \\
a_2 &= -0.253 \\
b_1 &= 2.411 \\
b_2 &= -0.29
\end{aligned}
\]
The corrected structural coefficient is then
\[
k_{\rm corr}(n) \;=\; \tfrac12\, A(n)\,\eta_{\rm fit}(n)
\]
which accurately reproduces the exact numerical values of $k(n)$
over the range $n=1$--$9$, while avoiding the strong bias of the
traditional approximation $k_{\rm approx}(n) = \tfrac12 A(n)$
\begin{table}[ht]
\centering
\caption{Exact, approximate, and corrected structural coefficients $k(n)$ at representative S\'ersic indices, together with the correction factor $\eta(n) = 2k_{\rm exact}/A(n)$.}
\begin{tabularx}{\textwidth}{c *{4}{>{\centering\arraybackslash}X}}
\hline
$n$ & $k_{\rm exact}$ & $k_{\rm approx}$ & $k_{\rm corr}$ & $\eta(n)$ \\
\hline
1 & 0.3141 & 0.5396 & 0.3154 & 0.584 \\
2 & 0.3092 & 0.7288 & 0.3089 & 0.424 \\
4 & 0.3373 & 1.3423 & 0.3364 & 0.251 \\
6 & 0.3909 & 2.4772 & 0.3933 & 0.159 \\
8 & 0.4662 & 4.5739 & 0.4657 & 0.102 \\
9 & 0.4997 & 6.2001 & 0.4989 & 0.080 \\
\hline
\end{tabularx}
\label{tab:k_values}
\end{table}

\subsection{Addressing the FP Tilt}
The observed FP of elliptical galaxies exhibits a tilt compared to the expectation from the simple virial theorem. This tilt can be understood through our derived scaling relation:
\begin{equation}
\frac{I_e R_e}{\sigma^2} = \frac{1 - f_{DM}}{2 \pi G \Upsilon_* k}
\end{equation}
where \( I_e \) is the effective surface brightness, \( R_e \) is the effective radius, \( \sigma \) is the velocity dispersion, \( f_{DM} \) is the dark matter fraction within \( R_e \), \( \Upsilon_* \) is the stellar mass-to-light ratio, \( G \) is the gravitational constant, and \( k \) is a structural constant dependent on the galaxy's light profile.
The inclusion of the dark matter fraction \( f_{DM} \), the stellar mass-to-light ratio \( \Upsilon_* \), and the structural constant \( k \) allows the relation to account for systematic variations across galaxies. For instance, more massive galaxies tend to have higher dark matter fractions, which affects the relationship between \( I_e \), \( R_e \), and \( \sigma \), leading to the observed tilt in the FP. Additionally, variations in \( \Upsilon_* \) due to differences in stellar populations and in \( k \) due to structural non-homology further contribute to this tilt.
Thus, our derived relation provides a theoretical framework that naturally incorporates the factors responsible for the FP tilt, offering a more comprehensive understanding of the dynamics and structure of elliptical galaxies.
\subsection{Derivation of the Virial Scaling Relation for Elliptical Galaxies Including Velocity Anisotropy}
The virial theorem for a steady, self-gravitating system gives
\begin{equation}
2K + W = 0
\end{equation}
As before, for an elliptical galaxy we write the kinetic energy in terms of the total mass \(M\) and a characteristic (observed) velocity dispersion \(\sigma_{\rm obs}\). To allow for velocity anisotropy we distinguish between the isotropic dispersion that enters the theoretical virial balance, \(\sigma_{\rm iso}\), and the observed (projected) dispersion \(\sigma_{\rm obs}\) which is affected by orbital anisotropy after line-of-sight projection. We therefore write
\begin{equation}
K = \frac{1}{2} M \sigma_{\rm iso}^2
\end{equation}
and the potential energy as
\begin{equation}
W = -\frac{G M^2}{R_e} k
\end{equation}
where \(k=k(n)\) encodes the structural (S\'ersic) non-homology as before.
From the virial theorem we obtain the isotropic virial relation
\begin{equation}
\sigma_{\rm iso}^2 = \frac{G M}{R_e} k \label{eq:sigma_iso}
\end{equation}
Following the Jeans-anisotropic (JAM) formalism \cite{Cappellari2008} for cylindrically-aligned anisotropy we define
\begin{equation}
\beta_z \equiv 1 - \frac{\sigma_z^2}{\sigma_R^2}
\end{equation}
and the projection introduces a multiplicative factor \(F(\beta_z,i)\) such that the \emph{observed} projected second moment (which we denote here \(\sigma_{\rm obs}^2\)) is related to the isotropic dispersion by
\begin{equation}
\sigma_{\rm obs}^2 = \sigma_{\rm iso}^2 \; F(\beta_z,i)
\end{equation}
with the JAM-motivated factor (for a cylindrically-aligned velocity ellipsoid)
\begin{equation}
F(\beta_z,i) \;=\; \cos^2 i \;+\; \frac{\sin^2 i}{1-\beta_z}
\label{eq:Fdef}
\end{equation}
where \(i\) is the inclination (angle between symmetry axis and line-of-sight; \(i=90^\circ\) is edge-on). Equation \eqref{eq:Fdef} follows directly from the projected second-moment expressions of the anisotropic Jeans solution.
Combining \eqref{eq:sigma_iso} and the projection relation gives a direct expression connecting the observed dispersion to the mass:
\begin{equation}
\sigma_{\rm obs}^2 \;=\; \left(\frac{G M}{R_e} k\right) F(\beta_z,i)
\end{equation}
Proceeding as in the isotropic case, split the total mass within \(R_e\) into stellar and dark components,
\begin{equation}
M = M_* + M_{\rm DM}, \qquad M_* = \Upsilon_* L
\end{equation}
and define the dark-matter fraction within \(R_e\),
\begin{equation}
f_{\rm DM} \equiv \frac{M_{\rm DM}}{M_* + M_{\rm DM}}
\end{equation}
Then
\begin{equation}
M = \frac{\Upsilon_* L}{1 - f_{\rm DM}}
\end{equation}
Substituting \(L = 2\pi I_e R_e^2\) and rearranging yields the anisotropy-inclusive virial scaling
\begin{equation}
\sigma_{\rm obs}^2
= \frac{2\pi G \Upsilon_* I_e R_e k}{1 - f_{\rm DM}} \; F(\beta_z,i)
\end{equation}
Equivalently, the FP style combination becomes
\begin{equation}
\;
\frac{I_e R_e}{\sigma_{\rm obs}^2}
\;=\;
\frac{1 - f_{\rm DM}}{2\pi G \Upsilon_*\, k\, F(\beta_z,i)}
\;
\label{eq:FP_aniso}
\end{equation}
\paragraph{Linearized approximation.}
For small anisotropy \((\beta_z \lesssim 0.2)\) it is often convenient to use the first-order expansion of \(F\) from \eqref{eq:Fdef}:
\begin{equation}
F(\beta_z,i) \simeq 1 + \beta_z \sin^2 i
\end{equation}
In that case
\begin{equation}
\frac{I_e R_e}{\sigma_{\rm obs}^2}
\approx
\frac{1 - f_{\rm DM}}{2\pi G \Upsilon_*\, k}\,\frac{1}{1+\beta_z\sin^2 i}
\approx
\frac{1 - f_{\rm DM}}{2\pi G \Upsilon_*\, k}\Big(1 - \beta_z\sin^2 i\Big)
\end{equation}
to first order in \(\beta_z\). Thus, to leading order anisotropy alters primarily the FP zero-point (and only weakly the slopes, unless \(\beta_z\) correlates systematically with \(I_e\) or \(R_e\)).
\paragraph{Spherical alternative.}
If one instead adopts the spherical Jeans approximation with anisotropy \(\beta\equiv1-\sigma_\theta^2/\sigma_r^2\), the projected second-moment integrals lead to an analogous rescaling; a compact approximate replacement is
\begin{equation}
\sigma_{\rm obs}^2 \simeq \sigma_{\rm iso}^2\,(1-\beta_{\rm eff})
\end{equation}
so that \eqref{eq:FP_aniso} becomes
\(\; \frac{I_e R_e}{\sigma_{\rm obs}^2} = \dfrac{1 - f_{\rm DM}}{2\pi G \Upsilon_*\, k\, (1-\beta_{\rm eff})}\;\)
where \(\beta_{\rm eff}\) is a suitably defined average anisotropy within the measurement aperture.
\paragraph{Remarks.}
The factor \(F(\beta_z,i)\) stems from the JAM projection formulae \cite{Cappellari2008}. It is appropriate for axisymmetric, cylindrically-aligned velocity ellipsoids (fast-rotator regime / ATLAS3D-like systems).
We may use the full \(F(\beta_z,i)\) (equation \ref{eq:Fdef}) if we have per-galaxy \(\beta_z\) and \(i\). If not, the linearized form is a defensible approximation for \(\beta_z\lesssim0.2\)
So The FP equation becomes
\begin{align}
\log R_e \;=\;&\; a\,\log \sigma 
+ b\,\log I_e 
+ d\,\log \Upsilon_\star \nonumber \\[4pt]
&+ e\,\log\!\bigl(1 - f_{\mathrm{DM}}\bigr) 
+ f\,\log k(n) \nonumber \\[4pt]
&+ g\,\log F(\beta_z,i) 
+ c
\end{align}
\section{Data Sample}
The dataset used for this analysis is based on the ATLAS³D survey, which includes 260 early-type galaxies (ETGs) observed in the local universe. The following parameters were extracted from the respective papers:
\textbf{Distance (D)}: The distances to the galaxies were taken from \cite{Cappellari2011a}, where the distance estimates (in Mpc) are provided.
\textbf{Effective radius}: The effective radii of the galaxies were retrieved from \cite{Krajnovic2013_ATLAS3D_XVII}, Table C1, where the values are given in arcseconds and converted to kpc using the distance estimates from \cite{Cappellari2011a}.
\textbf{Surface brightness ($\mu_e$)}: The mean surface brightness within $R_e$ was obtained from \cite{Krajnovic2013_ATLAS3D_XVII}, and converted to luminosity surface density (L$_{\odot}$/pc$^2$).
\textbf{Velocity dispersion ($\sigma_e$)}: The velocity dispersion within $R_e$ was sourced from \cite{Cappellari2013_XV}, which provides the luminosity-weighted velocity dispersion values for the ATLAS³D galaxies.
\textbf{Stellar mass-to-light ratio ($\Upsilon_*$) and Dark matter fraction ($f_{\text{DM}}$)}: These parameters were taken from \cite{Cappellari2013b}.
\textbf{Sérsic index ($n$)}: Sérsic index values for the galaxies were also extracted from\cite{Krajnovic2013_ATLAS3D_XVII}.
\textbf{Inclination ($i$)}: The inclination angle ($i$) was retrieved from \cite{Cappellari2013_XV}, where the inclination was derived from the dynamical models using the JAM fitting approach.
\textbf{Anisotropy ($\beta_z$)}: The anisotropy parameter \(\beta_z\) was taken from the electronic supplementary table which is available on the ATLAS3D project website \url{https://groups.physics.ox.ac.uk/atlas3d/}.
The distances (D) for the galaxies in the ATLAS3D sample were determined using several methods, each with different accuracies:
\textbf{Surface Brightness Fluctuation (SBF):}  
The distances were initially estimated using the Surface Brightness Fluctuation (SBF) method. The measurements were based on the ACS Virgo Cluster Survey, which provided distances with an uncertainty of approximately \textbf{3\%}.
\textbf{SBF from ground-based data:}  
Another set of distances was derived using ground-based SBF measurements from Tonry et al. (2001), which have an uncertainty of about \textbf{10\%}.
\textbf{NED-D (NASA Extragalactic Database):}  
Distances were also obtained from the NED-D catalog, which includes various methods such as Tip of the Red Giant Branch (TRGB), Cepheid variables, and Tully-Fisher Relation (TFR). The uncertainty for this method ranges between \textbf{10\% and 20\%}.
\textbf{Virgo Cluster distances:}  
For galaxies in the Virgo cluster, a fixed distance of \textbf{16.5 Mpc} was assumed, with an uncertainty of about \textbf{7\%}.
\textbf{Radial velocity-based distances:}  
For galaxies lacking direct distance measurements, distances were calculated from their heliocentric velocities (V$_{\text{hel}}$) using the local flow model. The uncertainty for these distances is around \textbf{21\%}.
So we took \textbf{10\%} uncertainty for the distance in this paper
\subsection{Dark matter fraction}
We adopt the values of the dark matter fraction within the effective radius,
$f_{\rm DM}(R_e)$, as provided by the ATLAS$^3$D project 
\cite{Cappellari2013b}. In their definition,
\begin{equation}
    f_{\rm DM}(R_e) \equiv \frac{M_{\rm DM}(<R_e)}{M_{\rm DM}(<R_e) + M_\star(<R_e)} \, 
\end{equation}
where the enclosed stellar mass $M_\star(<R_e)$ is derived from 
Multi-Gaussian Expansion photometry with a constant stellar mass-to-light 
ratio $\Upsilon_\star$, and the dark matter mass $M_{\rm DM}(<R_e)$ is 
computed from a parametric NFW halo model (``model B'' in Paper XV). 
The estimates are obtained from Jeans Anisotropic Modelling (JAM) fits to the 
stellar kinematics. In the ATLAS$^3$D, no individual error bars on 
$f_{\rm DM}(R_e)$ are given, but models with fit quality parameter 
${\rm qual}<2$ should be treated with caution. The typical random errors on 
related quantities are 5\% on $\sigma(R_e/8)$ and 6\% on the stellar 
mass-to-light ratio, which indirectly propagate into the uncertainty on 
$f_{\rm DM}(R_e)$. In this work we therefore regard $f_{\rm DM}(R_e)$ as an effective enclosed fraction within $R_e$, carrying systematic uncertainties at the level of several percent due to degeneracies between $\Upsilon_\star$ and the halo parameters. For this reason, we do not assign a formal error column to $f_{\rm DM}$ in our analysis, as the uncertainties are dominated by model systematics rather than direct observational errors.
\subsection{velocity dispersion and inclination angle}
In \cite{Cappellari2013_XV}, the effective stellar velocity dispersion $\sigma_e$ was measured by co-adding all SAURON spectra contained within the effective ellipse of area $A_e = \pi R_e^2$, defined by the effective radius $R_e$ and ellipticity $\epsilon_e$. A single combined spectrum was then analysed with the \texttt{pPXF} code (Cappellari \& Emsellem 2004), ensuring that $\sigma_e$ accounts for both the effects of stellar rotation and random motions. The associated random uncertainty on $\sigma_e$ is about \textbf{5\%} (corresponding to 0.021 dex).  
The inclination angle $i$ of each galaxy was obtained from the best-fitting mass-follows-light JAM (Jeans Anisotropic MGE) dynamical models. The error on $i$ is dominated by systematics and is not easy to quantify on an object-by-object basis, but for galaxies with reliable model fits the typical uncertainty is smaller than \textbf{$5^\circ$}.
\subsection{Stellar Population Mass-to-Light Ratio (\(\Upsilon_*\)).}
The stellar population mass-to-light ratio, \(\Upsilon_*\), was taken from\cite{Cappellari2013b}, where it was derived
via stellar population synthesis (SPS) modelling of the SAURON absorption
line-strength indices (H$\beta$, Fe5015, Mgb). The observed indices were fitted
with the MIUSCAT model library to infer the luminosity-weighted age and
metallicity of each galaxy, assuming a reference IMF. The corresponding
stellar mass-to-light ratio in the $r$-band, \(\Upsilon_*\), was then obtained
from these best-fit models. The tabulated values include the quoted
$1\sigma$ statistical uncertainties from the SPS fitting procedure, and in our
analysis we adopted these published errors directly.
\subsection{Sérsic Profile Parameters: \( \mu_e \), \( R_e \), and \( n \) with Uncertainties}
In this work we adopt the Sérsic profile parameters, namely the effective surface brightness 
\(\mu_e\), the effective radius \(R_e\) (in arcseconds), and the Sérsic index \(n\), as presented in 
\cite{Krajnovic2013_ATLAS3D_XVII}. These quantities were derived from two-dimensional photometric fits 
to the SDSS $r$-band images of the ATLAS$^{\rm 3D}$ galaxies, using the GALFIT 
algorithm applied to Multi-Gaussian Expansion (MGE) models. The effective surface brightness \(\mu_e\) corresponds to the mean brightness within the effective isophote, while the effective 
radius \(R_e\) represents the circularized half-light radius of the best-fitting Sérsic model. The Sérsic index \(n\) characterizes the concentration of the stellar light profile. The authors provide 
formal fitting uncertainties for these parameters, which are explicitly listed in the tables of the electronic supplement, and we take these errors into account in our analysis. In addition, we 
converted the effective radii from arcseconds into kiloparsecs, consistently propagating both the statistical fitting errors and a fixed $10\%$ uncertainty in the adopted galaxy distances \(D\).
\subsection{On the role of anisotropy}
The ATLAS$^{3D}$ team has explicitly cautioned that the fitted anisotropy 
parameters are only meaningful for galaxies with reliable dynamical models.
(i.e. inclination $>60^\circ$ and acceptable model quality). 
For the majority of objects in our full sample, these conditions are not met, 
and anisotropy values are therefore dominated by the inclination--anisotropy degeneracy. Including such quantities in a regression across the entire sample would add noise rather than astrophysically relevant signal.
For this reason, we did not include anisotropy in our full-sample fits 
(258 galaxies). Instead, we adopted a conservative approach: as a robustness check, we restricted to the ``OK'' subsample of 111 galaxies with reliable anisotropy estimates, and repeated the regression both with and without the anisotropy parameter. The inclusion of anisotropy did not improve the scatter 
or the overall $R^2$, and only produced a negligible shift in the zero-point 
($\sim 0.05$ dex).
We therefore conclude that anisotropy does not play a significant role in the FP relation at the level of precision explored here, and we avoid introducing spurious uncertainties by excluding unreliable anisotropy 
values from the full-sample fits.
\section{Results}
\begin{figure}[H]
    \centering
    \includegraphics[width=0.8\linewidth]{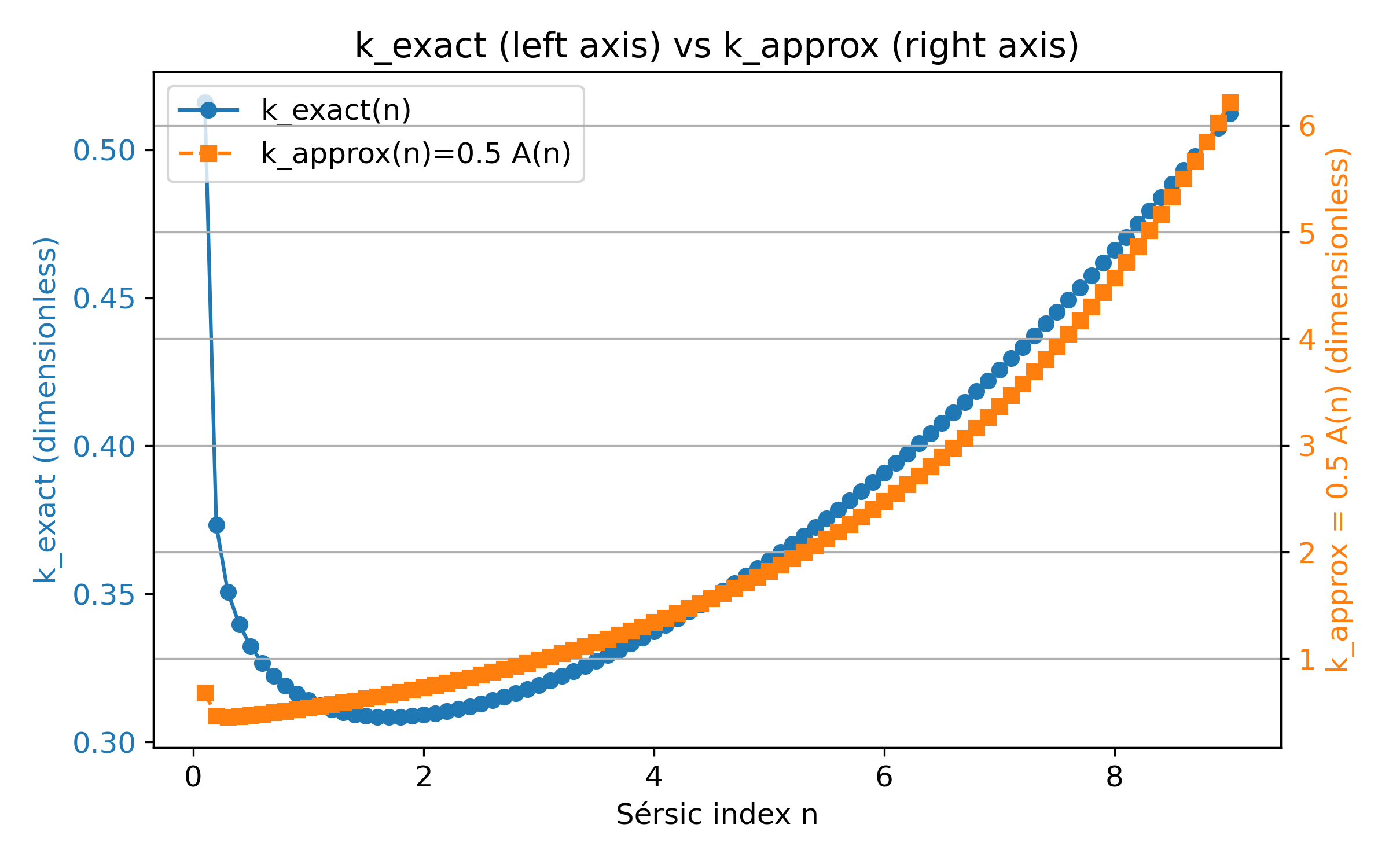}
    \caption{Comparison between the exact structural coefficient $k_{\rm exact}(n)$ and $k_{\rm approx}(n) = \tfrac{1}{2} A(n)$ as a function of the Sérsic index $n$
    }
    \label{fig:k_vs_approx}
\end{figure}
Figure~\ref{fig:k_vs_approx} illustrates the behavior of the exact structural coefficient $k_{\rm exact}(n)$ compared to the commonly used approximation $k_{\rm approx}(n) = 0.5\,A(n)$ as a function of the Sérsic index $n$. The exact coefficient, computed from the fundamental integral relation involving the mass and potential distribution, shows a relatively mild increase from $n=1$ to $n\approx 9$, ranging from $k_{\rm exact}\sim 0.31$ to $\sim 0.50$. In contrast, the approximate coefficient rises dramatically with $n$, exceeding $6$ at $n=9$. This divergence highlights the strong bias introduced by the assumption $W \approx \tfrac{1}{2} M \Phi_0$, which increasingly overestimates the true potential energy for highly concentrated systems. Consequently, for accurate modeling of elliptical galaxies with large Sérsic indices, the use of $k_{\rm exact}(n)$ or a corrected form calibrated to it is essential to avoid systematic errors in dynamical mass estimates and FP analyses.
\begin{figure}[H]
    \centering
    \includegraphics[width=0.8\linewidth]{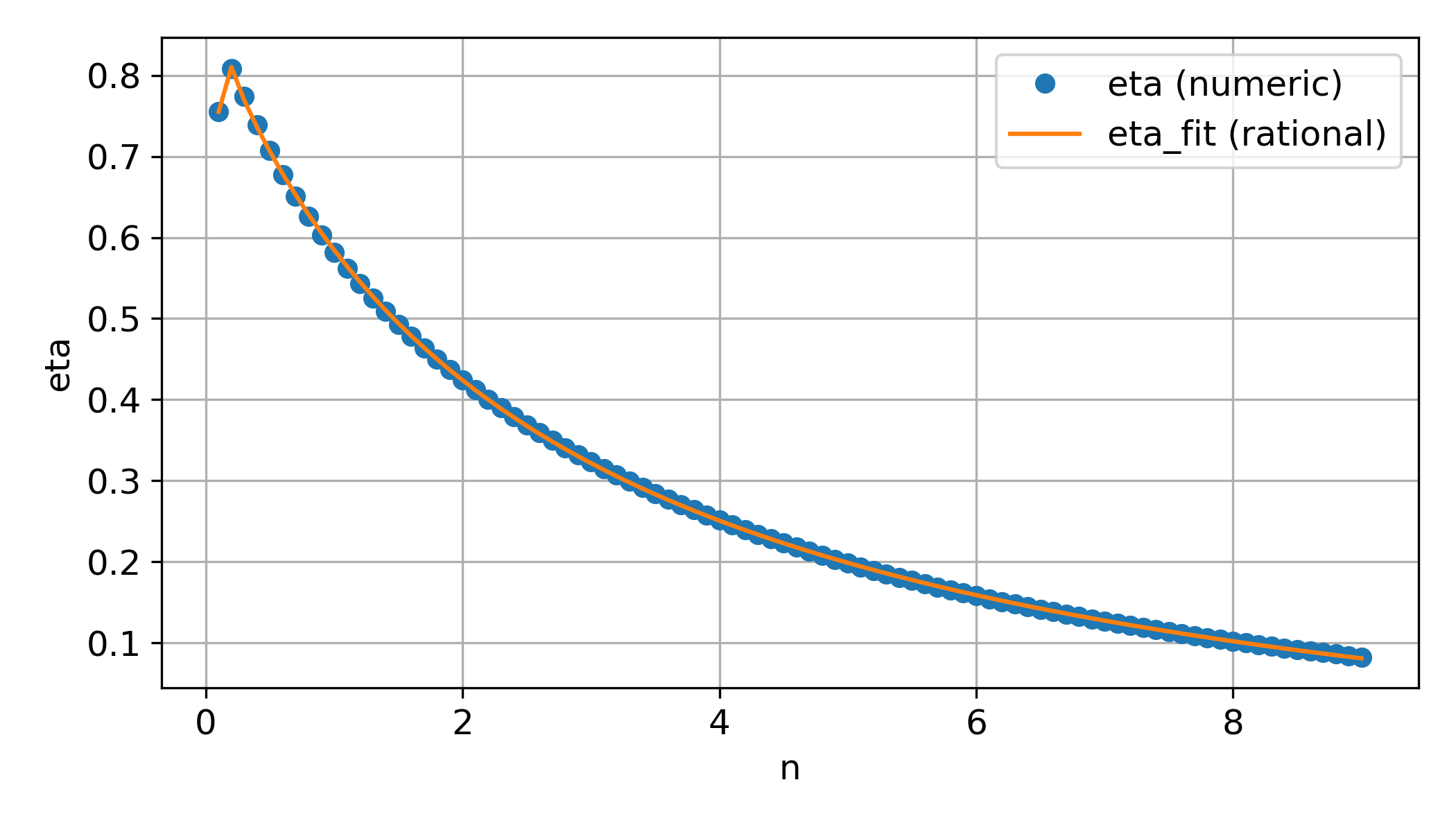}
    \caption{Numerical values of $\eta(n)$ compared with the rational fitting formula}
    \label{fig:eta_fit}
\end{figure}
Figure~\ref{fig:eta_fit} shows the correction factor $\eta(n) = \langle \Phi \rangle / \Phi_0$ as a function of the Sérsic index $n$, comparing the numerically computed values (blue points) with the rational fitting function $\eta_{\rm fit}(n)$ (orange curve). The correction factor quantifies the departure from the simplified approximation $W \approx \tfrac{1}{2} M \Phi_0$, with $\eta(n) \le 1$ for all $n$. The numerical results demonstrate a strong monotonic decline from $\eta \approx 0.8$ at $n \approx 0.5$ to below $0.1$ at $n \approx 9$, indicating that the approximation increasingly overestimates the potential energy for highly concentrated profiles. The fitted expression provides an accurate analytic representation of the numerical trend, ensuring a practical correction for use in dynamical modeling and FP analyses.
We compared the classical FP relation:
\begin{equation}
\log R_e = a \log \sigma + b \log \langle I_e \rangle + c
\end{equation}
with an extended formulation that incorporates additional physical parameters:
\begin{equation} \log R_e = a\,\log \sigma + b\,\log I_e + d\,\log \Upsilon_\star + e\,\log(1 - f_{\mathrm{DM}}) + f\,\log k(n) + c \end{equation}
and \begin{align}
\log R_e \;=\;&\; a\,\log \sigma 
+ b\,\log I_e 
+ d\,\log \Upsilon_\star \nonumber \\[4pt]
&+ e\,\log\!\bigl(1 - f_{\mathrm{DM}}\bigr) 
+ f\,\log k(n) \nonumber \\[4pt]
&+ g\,\log F(\beta_z,i) 
+ c
\end{align}
\subsection{FP Fits and Robustness Checks}
We present here the full results of the regression fits for both the classical and extended 
 FP, together with additional robustness checks (multicollinearity and bootstrap resampling).
\subsubsection{Classical FP fits}
\textbf{OLS, full sample (n = 258):} \\
    const $= -9.45 \pm 0.32$, \ 
    $\log \sigma = 0.760 \pm 0.055$, \ 
    $\log I_e = -0.546 \pm 0.019$; \\
    $R^2 = 0.793$, \ AIC/BIC = $-273.3/-262.6$, \ RMS scatter = $0.141$ dex.
\textbf{WLS, full sample (n = 258):} \\
    const $= -9.58 \pm 0.38$, \ 
    $\log \sigma = 0.773 \pm 0.051$, \ 
    $\log I_e = -0.554 \pm 0.024$; \\
    $R^2 = 0.736$, \ AIC/BIC = $-211.7/-201.1$, \ RMS scatter = $0.142$ dex.
\subsubsection{Extended FP fits}
\textbf{OLS, full sample (n = 258):} \\
    const $= -13.17 \pm 0.48$, \ 
    $\log \sigma = 1.243 \pm 0.079$, \ 
    $\log I_e = -0.693 \pm 0.023$, \\
    $\log (M/L) = -0.433 \pm 0.068$, \ 
    $\log (1-f_{\rm DM}) = 0.479 \pm 0.080$, \ 
    $\log k(n) = -1.582 \pm 0.192$; \\
    $R^2 = 0.848$, \ AIC/BIC = $-347.3/-326.0$, \ RMS scatter = $0.121$ dex.
\textbf{WLS, full sample (n = 258):} \\
    const $= -12.88 \pm 0.58$, \ 
    $\log \sigma = 1.200 \pm 0.081$, \ 
    $\log I_e = -0.674 \pm 0.027$, \\
    $\log (M/L) = -0.407 \pm 0.070$, \ 
    $\log (1-f_{\rm DM}) = 0.505 \pm 0.083$, \ 
    $\log k(n) = -1.742 \pm 0.287$; \\
    $R^2 = 0.787$, \ AIC/BIC = $-260.6/-239.3$, \ RMS scatter = $0.122$ dex.
\textbf{WLS, restricted OK\_sub sample, no anisotropy (n = 111):} \\
    const $= -13.76 \pm 0.99$, \ 
    $\log \sigma = 1.152 \pm 0.144$, \ 
    $\log I_e = -0.731 \pm 0.046$, \\
    $\log (M/L) = -0.490 \pm 0.123$, \ 
    $\log (1-f_{\rm DM}) = 0.512 \pm 0.210$, \ 
    $\log k(n) = -2.179 \pm 0.491$; \\
    $R^2 = 0.781$, \ AIC/BIC = $-115.3/-99.0$, \ RMS scatter = $0.128$ dex
\textbf{WLS, restricted OK\_sub sample, with anisotropy (n = 111):} \\
    const $= -13.80 \pm 0.99$, \ 
    $\log \sigma = 1.168 \pm 0.147$, \ 
    $\log I_e = -0.731 \pm 0.046$, \\
    $\log (M/L) = -0.499 \pm 0.124$, \ 
    $\log (1-f_{\rm DM}) = 0.530 \pm 0.214$, \ 
    $\log k(n) = -2.193 \pm 0.493$, \\
    $\log F = 0.074 \pm 0.137$; \\
    $R^2 = 0.782$, \ AIC/BIC = $-113.6/-94.6$, \ RMS scatter = $0.128$ dex
\subsubsection{Zero-point comparison}
\noindent
Extended (WLS, full): $-12.88$; \\
Extended (WLS, OK\_sub, no anisotropy): $-13.76$; \\
Extended (WLS, OK\_sub, + anisotropy): $-13.80$; \\
Shift due to anisotropy (OK\_sub): $\Delta \approx -0.046$ dex
\subsubsection{Robustness checks}
\paragraph{Variance Inflation Factors (VIF).} 
For the extended FP (full sample), the VIF values are:
\[
\log (M/L): 3.96, \quad 
\log \sigma: 2.82, \quad 
\log (1-f_{\rm DM}): 2.14, \quad 
\log I_e: 1.95, \quad 
\log k(n): 1.86.
\]
All are below the conservative threshold of 5, indicating no severe multicollinearity.
\paragraph{Bootstrap resampling.}
Using 2000 bootstrap realizations of the extended FP (full sample), 
we obtain the following coefficient distributions (mean, standard deviation, and 95\% confidence intervals):

\begin{center}
\begin{tabular}{lcccc}
\hline
Parameter & Mean & Std & 2.5\% & 97.5\% \\
\hline
const & $-13.17$ & 0.50 & $-14.16$ & $-12.15$ \\
$\log \sigma$ & $1.246$ & 0.087 & $1.081$ & $1.416$ \\
$\log I_e$ & $-0.693$ & 0.023 & $-0.738$ & $-0.646$ \\
$\log (M/L)$ & $-0.434$ & 0.080 & $-0.589$ & $-0.275$ \\
$\log (1-f_{\rm DM})$ & $0.478$ & 0.086 & $0.316$ & $0.650$ \\
$\log k(n)$ & $-1.586$ & 0.174 & $-1.938$ & $-1.251$ \\
\hline
\end{tabular}
\end{center}

All bootstrap confidence intervals exclude zero, confirming the robustness and statistical significance of the additional predictors.
\subsubsection{Statistical Comparisons between Classical and Extended FP}
To quantitatively assess the improvement of the extended FP over the classical formulation, 
we performed a series of statistical model comparison tests:
\textbf{ANOVA (nested F-test).}  
    Comparing the classical FP against the extended FP yields  
    $F = 30.53$ with $p \approx 7.2 \times 10^{-17}$,  
    confirming that the additional predictors significantly improve the fit.
\textbf{Adjusted $R^2$}  
    Classical: $\mathrm{adj}\,R^2 = 0.791$,  
    Extended: $\mathrm{adj}\,R^2 = 0.845$,  
    with $\Delta \mathrm{adj}\,R^2 \approx 0.054$.
\textbf{AIC and Akaike weights.}  
    AIC values: Classical = $-273.3$, Extended = $-347.3$.  
    Akaike weights strongly favour the extended FP  
    (classical $\approx 8.6 \times 10^{-17}$, extended $\approx 1.0$).
\textbf{Cross-validation (5-fold).}  
    Mean RMSE: Classical $= 0.141$ dex, Extended $= 0.123$ dex.  
    Improvement: $\Delta \mathrm{RMSE} \approx -0.018$ dex.
\textbf{Paired bootstrap (RMSE difference).}  
    Mean difference $= -0.0208$ dex, but empirical $p \approx 0.48$,  
    indicating that while the trend favours the extended FP,  
    the improvement is not always significant across bootstrap realizations.
\textbf{Residual diagnostics.}  
    Kolmogorov–Smirnov test: $p = 0.974$ (no departure from normality).  
    Levene’s test: $p = 0.068$ (no strong evidence for heteroscedasticity).
\textbf{Partial $R^2$ of added predictors.}  
    The additional predictors in the extended FP contribute  
    a partial $R^2 \approx 0.36$, i.e. they explain about 36\% of the residual variance 
    left over from the classical FP.
\begin{table}[ht]
\centering
\caption{Statistical significance of the regression coefficients from the weighted least-squares (WLS) fit of the extended FP. Reported values are the p-values associated with each parameter.}
\label{tab:wls_pvalues}
\begin{tabular}{|l|c|}
\hline
\textbf{Variable} & \textbf{p-value (WLS)} \\
\hline
Const.        & $< 10^{-16}$ \\
$\log \sigma$ & $< 10^{-16}$ \\
$\log I_e$    & $< 10^{-16}$ \\
$\log (M/L)$  & $1.59 \times 10^{-8}$ \\
$\log (1-f_{\rm DM})$ & $4.21 \times 10^{-9}$ \\
$\log k(n)$   & $4.79 \times 10^{-9}$ \\
\hline
\end{tabular}
\end{table}
The weighted least-squares regression indicates that all coefficients in the extended FP relation are highly statistically significant. 
The structural parameters ($\log \sigma$, $\log I_e$) show extremely strong significance ($p < 10^{-16}$), confirming their central role in defining the Plane. The additional terms accounting for stellar mass-to-light ratio ($\log M/L$), dark matter fraction ($\log (1-f_{\rm DM})$), and the structural coefficient $k(n)$ are also significant at the $p < 10^{-8}$ level, demonstrating that they provide robust, non-negligible contributions to the regression.
\subsection{RANSAC Robustness Test}
To check the effect of outliers on the FP fits, we also applied the RANSAC algorithm. This method repeatedly fits the model to random subsamples of the data and keeps only the inliers that are consistent with the relation, 
while rejecting strong outliers (e.g.\ cD galaxies and dwarfs). 
This provides a simple way to test whether the results of the extended FP remain stable when the sample contains heterogeneous systems. Using the RANSAC algorithm on the full sample of 258 galaxies. This approach iteratively identifies inliers consistent with the model while rejecting strong outliers (e.g. dwarfs and cD galaxies).
For the \textbf{classical FP} under RANSAC (WLS), we find:
\[
R^2 = 0.926, \quad \mathrm{RMS} = 0.142~\mathrm{dex}, \quad \mathrm{inliers} = 164/258
\]
For the \textbf{extended FP}, we obtain:
\[
R^2 = 0.949, \quad \mathrm{RMS} = 0.124~\mathrm{dex}, \quad \mathrm{inliers} = 173/258
\]
Thus, even in the presence of the full heterogeneous sample, including dwarf and cD galaxies, the extended FP achieves both a significantly higher goodness-of-fit and a larger number of inliers. The improvement in scatter is
\[
\Delta \mathrm{RMS} = -0.017~\mathrm{dex}
\]
and the $R^2$ increases by $\Delta R^2 \simeq 0.024$ compared to the classical FP. This demonstrates that the improvement of the extended FP is not an artifact of restricting the sample, but reflects genuine additional explanatory power provided by the stellar $M/L$, the dark matter fraction, and the structural coefficient $k(n)$.
For the high-quality ATLAS$^{\rm 3D}$ subset ($n=111$, $i > 60^\circ$, quality $>0$), the results are consistent. The extended FP yields $R^2 \simeq 0.949$ with or without including anisotropy, while the classical FP remains less predictive. Moreover, the inclusion of the anisotropy parameter has negligible impact on the scatter, confirming our earlier conclusion that anisotropy plays only a secondary role in setting the FP zero-point.
\begin{figure}[H]
    \centering
    \includegraphics[width=0.8\linewidth]{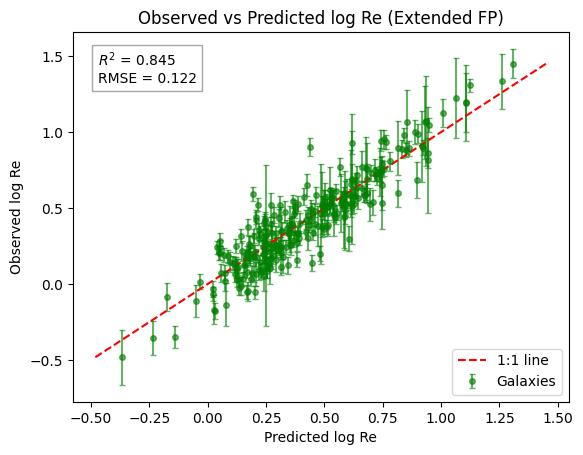}
    \caption{Comparison of observed effective radii (log $R_{e}$) against predicted values from the extended Fundamental Plane (FP) model}
    \label{fig:extfp_fit}
\end{figure}
Figure~\ref{fig:extfp_fit} demonstrates that the inclusion of structural and stellar population parameters leads to a substantially more accurate calibration. The points cluster tightly around the one-to-one line, indicating that the extended formulation captures most of the intrinsic variance in galaxy sizes. The reduced scatter implies that the residuals are dominated by observational uncertainties rather than systematic model deficiencies. This strong agreement suggests that the Extended FP represents a physically calibrated scaling relation where structural non-homology and stellar populations are explicitly taken into account. In turn, this provides a more reliable predictor of galaxy sizes across the full sample. At the same time, galaxies at the extreme ends show systematic deviations: a handful of systems with very large observed radii lie above the one-to-one line, consistent with cD-like or cluster-dominant galaxies whose extended stellar envelopes inflate the effective radius beyond the prediction. Conversely, several compact ellipticals and low-$n$ systems fall below the line, where the model slightly overpredicts $R_e$. These edge cases highlight the role of environmental effects and structural peculiarities (cD halos, compactness) as secondary drivers of the scatter around the Extended FP.

\begin{figure}[H]
    \centering
    \includegraphics[width=0.8\linewidth]{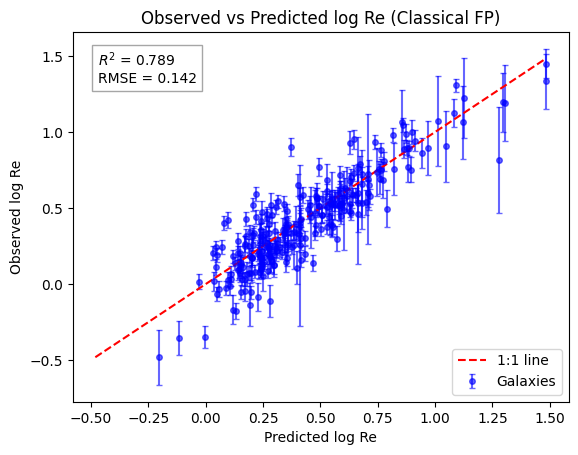}
    \caption{Comparison of observed log $R_{e}$ against predicted values from the classical Fundamental Plane (FP) model}
    \label{fig:classfp_fit}
\end{figure}
Figure ~\ref{fig:classfp_fit} shows a weaker performance relative to the extended formulation. While the overall correlation is evident, the scatter around the one-to-one line is noticeably larger, particularly for intermediate-size systems. This indicates that the classical FP does not fully capture structural or stellar population variations. Galaxies at the extremes highlight these shortcomings: several extended systems lie above the line, where the observed $R_e$ is larger than predicted, while compact ellipticals and low-mass systems fall significantly below the line. These deviations suggest that neglecting non-homology and dark matter effects reduces the predictive power of the classical FP and leaves systematic residuals at both ends of the galaxy population. 

\begin{figure}[H]
    \centering
    \includegraphics[width=0.8\linewidth]{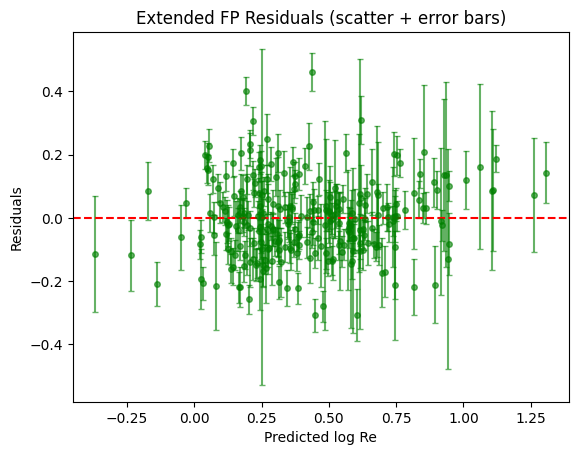}
    \caption{Distribution of residuals for the extended FP model }
    \label{fig:extfp_res}
\end{figure}
Figure~\ref{fig:extfp_res} provides further support for its improved performance. The deviations are symmetrically distributed around zero, with no apparent trends or systematic biases across the galaxy population. The residual standard deviation of 0.1221 and reduced $\chi^2$ of 4.65 confirm that the scatter is substantially lower than in the classical case. The lack of systematic structure in the residuals indicates that the Extended FP has successfully incorporated the primary physical drivers of galaxy size, leaving only stochastic scatter. This statistical robustness strengthens the case for the Extended FP as a more fundamental representation of elliptical galaxy scaling relations.

\begin{figure}[H]
    \centering
    \includegraphics[width=0.6\linewidth]{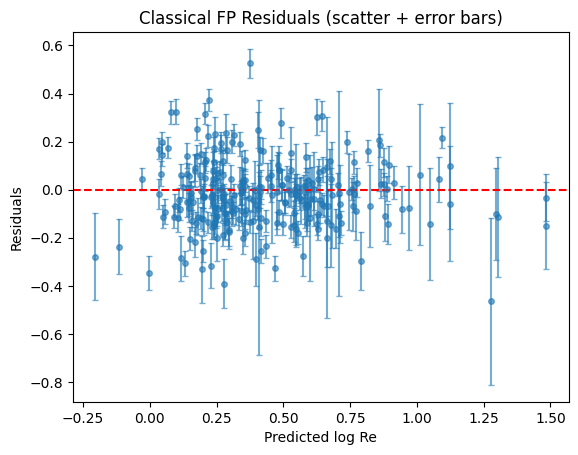}
    \caption{Distribution of residuals for the classical FP model}
    \label{fig:classfp_res}
\end{figure}
In contrast, Figure~\ref{fig:classfp_res} the residuals from the Classical FP reveal a broader and less symmetric distribution. A standard deviation of 0.1412 and a reduced $\chi^2$ of 5.75 confirm the weaker predictive power of the traditional formulation. The distribution shows evidence of mild systematic shifts, suggesting that important physical effects are not adequately modeled. This underlines the limitations of assuming strict homology and neglecting stellar population variations. The contrast between Figures \ref{fig:extfp_res} and \ref{fig:classfp_res} illustrates how the Extended FP achieves a significant reduction in residual variance.

\begin{figure}[H]
    \centering
    \includegraphics[width=0.6\linewidth]{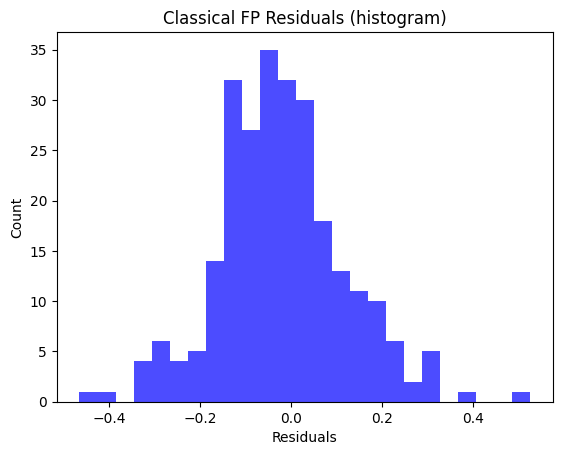}
    \caption{Histogram of residuals from the classical FP model}
    \label{fig:classfp_histo}
\end{figure}
Figure~\ref{fig:classfp_histo} illustrates a wide distribution with extended wings, reflecting both higher variance and potential unmodeled systematics. The reduced $\chi^2$ of 5.75 quantitatively confirms the poor statistical fit. The long tails of the distribution suggest that a subset of galaxies is systematically misrepresented by the classical formulation. This indicates that the simplifying assumptions of the traditional FP are not appropriate for the observed diversity of galaxy structures. The histogram therefore provides visual and statistical evidence of the inadequacy of the classical model.

\begin{figure}[H]
    \centering
    \includegraphics[width=0.6\linewidth]{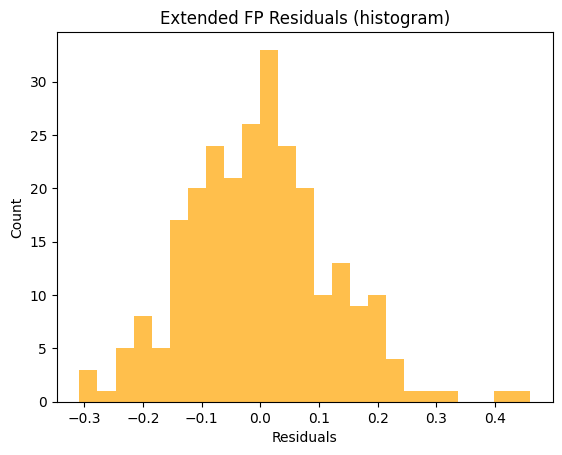}
    \caption{Histogram of residuals from the extended FP model}
    \label{fig:extfp_histo}
\end{figure}
In contrast, figure~\ref{fig:extfp_histo} displays a narrow distribution concentrated around zero. With a standard deviation of 0.1221 and reduced $\chi^2$ of 4.65, the fit is clearly superior to that of the classical case. The absence of strong wings in the histogram suggests that the extended formulation captures the behavior of both low- and high-Sérsic index galaxies without leaving large systematic outliers. This indicates that the Extended FP offers a more universal description of elliptical galaxy scaling, with residuals consistent with random scatter rather than systematic failure. Such behavior is essential for establishing the Extended FP as a reliable physical relation.
\begin{figure}[H]
    \centering
    \includegraphics[width=0.6\linewidth]{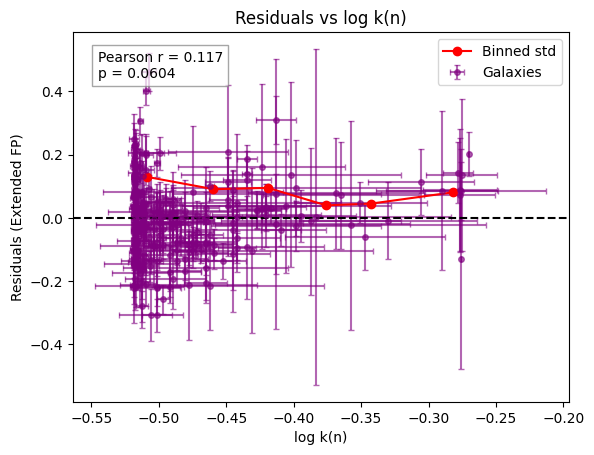}
    \caption{Residuals of the extended Fundamental Plane (FP) against the Sérsic structure parameter log $k(n)$}
    \label{fig:res_kn}
\end{figure}
Figure~\ref{fig:res_kn} of the residuals of the Extended FP are examined as a function of the S\'ersic structure parameter $k(n)$, with measurement uncertainties explicitly included on both axes. Error bars, particularly those associated with $n$, are substantial and contribute to the vertical scatter. After accounting for these uncertainties, only a weak correlation is detected ($r=0.117$, $p=0.060$), which does not reach formal statistical significance. To further assess possible systematic effects, the residuals were binned in $\log k(n)$, and the standard deviation within each bin was calculated (red points). The binned trend shows a mild increase in scatter at higher $k(n)$ (i.e., larger Sérsic index $n$), but the effect remains modest compared to the overall distribution. This indicates that most of the structural dependence is already absorbed by the Extended FP calibration, leaving only marginal residual variations. The inclusion of error bars demonstrates that part of the apparent trend may stem from measurement uncertainties

\begin{figure}[H]
    \centering
    \includegraphics[width=0.6\linewidth]{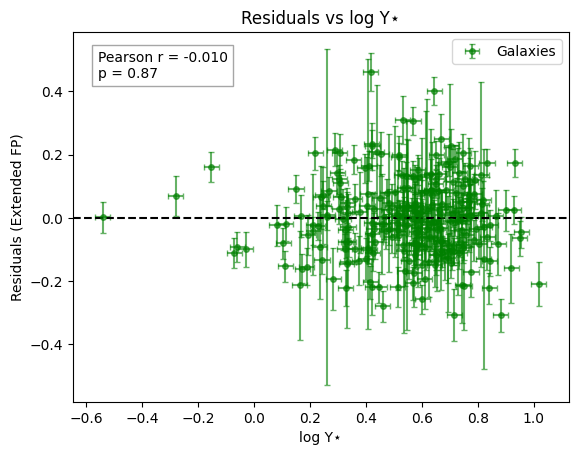}
    \caption{Residuals of the extended Fundamental Plane (FP) against the stellar mass-to-light ratio (log $\Upsilon_{*}$)}
    \label{fig:res_ml}
\end{figure}
Figure~\ref{fig:res_ml} of the residuals of the Extended FP are examined as a function of the stellar mass-to-light ratio $\Upsilon_{\star}$, with measurement uncertainties shown for both axes. The error bars demonstrate that observational uncertainties dominate much of the vertical scatter. The Pearson correlation test yields $r=-0.010$ with $p=0.87$, indicating no statistically significant dependence. This confirms that stellar population variations in age and metallicity have been effectively absorbed by the extended calibration. The absence of a trend, even after including error bars, implies that population-driven differences across the sample are not a dominant source of scatter in the relation.

\begin{figure}[H]
    \centering
    \includegraphics[width=0.6\linewidth]{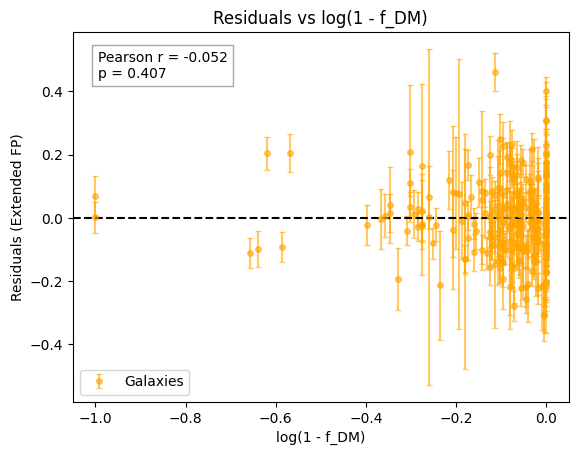}
    \caption{Residuals of the extended Fundamental Plane (FP) versus the central dark-matter fraction (log(1$-f_{\rm DM}$))}
    \label{fig:res_fdm}
\end{figure}
Figure ~\ref{fig:res_fdm} of the residuals of the Extended FP are compared to the central dark-matter fraction to test whether variations in baryonic versus dynamical contributions affect the scatter. Measurement uncertainties are included, and the error bars illustrate that observational errors contribute substantially to the vertical spread. The Pearson correlation test yields $r=-0.052$ with $p=0.40$, indicating no statistically significant linear dependence. However, the scatter visibly increases toward higher dark-matter fractions, suggesting potential limitations in the assumed dark-matter scaling within $R_{e}$. This behavior implies that, while the Extended FP is largely governed by baryonic structural parameters, residual deviations at high $f_{\mathrm{DM}}$ may reflect unmodeled complexities in halo structure or baryon–dark matter coupling. Overall, the lack of a significant correlation supports the interpretation that dark-matter fraction is not the dominant driver of the FP scatter
\begin{figure}[H]
    \centering
    \includegraphics[width=0.6\linewidth]{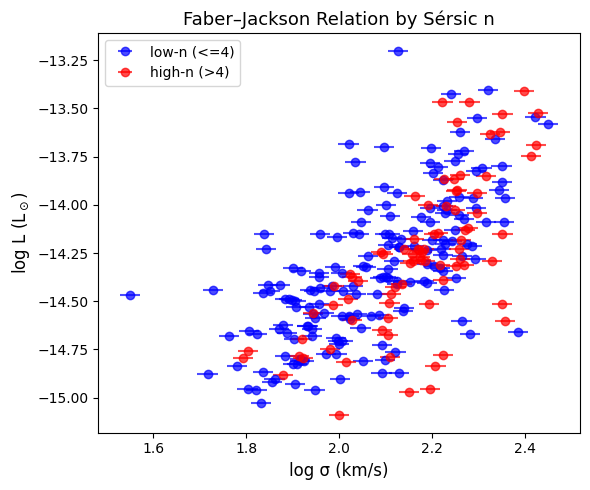}
    \caption{Faber-Jackson relation}
    \label{fig:faber_jackson}
\end{figure}
Figure ~\ref{fig:faber_jackson} of the classical Faber-Jackson relation is recovered for the sample, confirming the consistency of the data set with established luminosity–velocity dispersion scaling. Although scatter is present, the relation remains well defined, with brighter galaxies exhibiting higher velocity dispersions. The agreement between this empirical law and the Extended FP predictions ensures that the new calibration does not violate classical scaling relations. Instead, the Extended FP generalizes these laws by embedding them in a physically motivated framework that accounts for additional parameters.

\begin{figure}[H]
    \centering
    \includegraphics[width=0.6\linewidth]{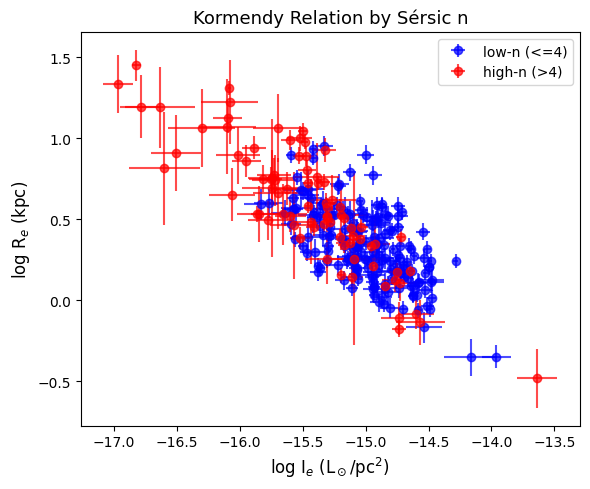}
    \caption{Kormendy relation}
    \label{fig:kormendy}
\end{figure}
As shown in Figure~\ref{fig:kormendy}, the Kormendy relation, linking surface brightness to effective radius, is shown to depend significantly on S\'ersic index. An ANCOVA analysis demonstrates that high-$n$ galaxies follow a steep slope of $-0.567$, while low-$n$ systems deviate, producing a measurable interaction effect ($p=0.027$). This implies that structural non-homology directly influences the Kormendy relation, and a single slope cannot adequately describe the full population. The Extended FP therefore provides a more unified framework by explicitly including structural parameters that explain these deviations. This result reinforces the interpretation that structural diversity must be incorporated in fundamental scaling laws.
\begin{figure}[H]
    \centering
    \includegraphics[width=0.7\linewidth]{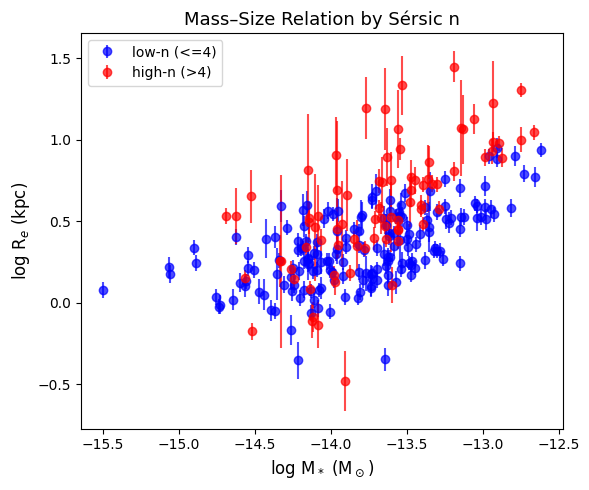}
    \caption{Mass-size relation}
    \label{fig:mass_size}
\end{figure}
As shown in Figure~\ref{fig:mass_size}, the mass–size relation shows a similar dependence on Sérsic index. ANCOVA analysis reveals that low-$n$ galaxies follow a significantly shallower slope, with an interaction term of $-0.25$ ($p<0.001$), while high-$n$ galaxies retain a steeper relation. This demonstrates that structural non-homology systematically alters the scaling between galaxy mass and size. The Extended FP accommodates this effect by embedding Sérsic-dependent terms, thereby offering a unified description across the full range of ellipticals. The results confirm that non-homology is not a marginal effect but a key factor shaping the scaling relations of galaxies.
\subsection{Spearman Correlation: \(\log k(n)\) vs \(n\)}
The Spearman rank correlation between \(\log k(n)\) and the Sérsic index \(n\) is found to be \(\rho = 0.939\) with a highly significant p-value of \(4.547 \times 10^{-121}\). This strong positive correlation indicates a near-monotonic relationship between galaxy concentration (represented by \(k(n)\)) and the Sérsic index. In other words, as the Sérsic index increases, galaxies tend to become more concentrated in their light distribution. This result is particularly important in the context of the Extended FP, as it suggests that structural parameters such as Sérsic index and concentration are closely tied. The high correlation further supports the validity of the Extended FP as a more accurate model, integrating these key structural variables to better describe the scaling of elliptical galaxies. This finding aligns with the notion that S\'ersic index captures essential structural variations that influence the effective radius, reaffirming the need to account for these factors in fundamental scaling relations. The robustness of the Spearman correlation provides a strong quantitative basis for the inclusion of Sérsic index in the Extended FP framework.
\subsection{de Vaucouleurs profile}
\begin{table}[htbp]
\centering
\caption{P-values from WLS regression for the full galaxy sample and the $n \approx 4$ subsample.}
\label{tab:pvals}
\begin{tabular}{|l|c|c|}
\hline
\textbf{Variable} & \textbf{Full sample (N=258)} & \textbf{$n \approx 4$ subsample (N=44)} \\
\hline
const      & $< 1 \times 10^{-16}$ & $2.38 \times 10^{-13}$ \\
$\log \sigma$   & $< 1 \times 10^{-16}$ & $4.57 \times 10^{-9}$ \\
$\log I_e$      & $< 1 \times 10^{-16}$ & $< 1 \times 10^{-16}$ \\
$\log (M/L)_\star$ & $1.59 \times 10^{-8}$ & $2.90 \times 10^{-3}$ \\
$\log(1-f_{\rm DM})$ & $4.21 \times 10^{-9}$ & $1.11 \times 10^{-1}$ \\
$\log k(n)$     & $4.79 \times 10^{-9}$ & $5.83 \times 10^{-2}$ \\
\hline
\end{tabular}
\end{table}

\begin{table}[htbp]
\centering
\caption{Summary metrics for the WLS regression fits.}
\label{tab:metrics}
\begin{tabular}{|l|c|c|}
\hline
 & \textbf{Full sample} & \textbf{$n \approx 4$ subsample} \\
\hline
$R^2$ & 0.787 & 0.909 \\
AIC   & -260.61 & -83.31 \\
BIC   & -239.29 & -72.61 \\
\hline
\end{tabular}
\end{table}

\paragraph{Interpretation.}
For the full sample ($N=258$), all coefficients are highly significant ($p < 10^{-8}$), indicating that 
velocity dispersion, surface brightness, stellar mass-to-light ratio, dark matter fraction, and structural coefficient $k(n)$ all contribute to the FP relation. In contrast, in the restricted $n \approx 4$ subsample ($N=44$), only $\log \sigma$, $\log I_e$, and $\log (M/L)_\star$ remain statistically significant. 
Both $\log(1-f_{\rm DM})$ and $\log k(n)$ lose significance ($p > 0.05$), suggesting that once the S\'ersic index is fixed near the \cite{deVaucouleurs1948} value, the FP reduces to its classical dependence on kinematics and photometric parameters. This comparison highlights the importance of structural diversity (via $n$ and $k(n)$) and dark matter contributions in the global FP, while confirming that within a homogeneous \cite{deVaucouleurs1948}-like subsample the FP is primarily governed by stellar dynamics and surface brightness. 
\subsection{Test of Removing Dark Matter Fraction}
\begin{table}[htbp]
\centering
\caption{Comparison of regression coefficients and p-values with and without $\log(1-f_{\rm DM})$.}
\begin{tabular}{|l|c|c|c|c|}
\hline
Variable & Coef (with) & p (with) & Coef (without) & p (without) \\
\hline
const       & -12.88 & $<1\times10^{-16}$ & -11.45 & $<1\times10^{-16}$ \\
$\log \sigma$ & $1.200$ & $<1\times10^{-16}$ & $0.975$ & $1.46\times10^{-28}$ \\
$\log I_e$   & $-0.674$ & $<1\times10^{-16}$ & $-0.612$ & $1.11\times10^{-63}$ \\
$\log (M/L)$ & $-0.407$ & $1.59\times10^{-8}$ & $-0.104$ & $4.68\times10^{-2}$ \\
$\log (1-f_{\rm DM})$ & $0.506$ & $4.21\times10^{-9}$ & --- & --- \\
$\log k$     & $-1.742$ & $4.79\times10^{-9}$ & $-1.265$ & $2.64\times10^{-5}$ \\
\hline
\end{tabular}
\end{table}

\begin{table}[htbp]
\centering
\caption{Model metrics with and without $\log(1-f_{\rm DM})$.}
\begin{tabular}{|l|c|c|c|}
\hline
Model & $R^2$ & AIC & BIC \\
\hline
With $\log(1-f_{\rm DM})$ & 0.787 & -260.61 & -239.29 \\
Without $\log(1-f_{\rm DM})$ & 0.755 & -227.19 & -209.43 \\
\hline
\end{tabular}

\vspace{0.3cm}
Likelihood-ratio test: $\chi^2 = 35.42$, $df=1$, $p = 2.7\times10^{-9}$.
\end{table}
\paragraph{Interpretation.}
Including the dark matter fraction term, $\log(1-f_{\rm DM})$, significantly improves the fit of the FP relation. 
The model with this variable achieves a higher $R^2$ and substantially lower AIC/BIC compared to the restricted model. 
The likelihood-ratio test confirms that the improvement is highly significant ($p \ll 0.001$).
Moreover, excluding $\log(1-f_{\rm DM})$ induces noticeable changes in the estimated coefficients: 
most prominently, the $\log(M/L)$ term loses its strong statistical significance (from $p \sim 10^{-8}$ to marginal $p \sim 0.05$), 
and the coefficients of $\log \sigma$ and $\log k$ are shifted toward weaker values. 
This indicates that $\log(1-f_{\rm DM})$ captures essential physical information related to the interplay between stellar populations, 
dynamical structure, and dark matter content. In summary, the inclusion of the dark matter fraction is statistically indispensable and physically meaningful for a robust calibration of the extended FP.
\subsection{Role of velocity dispersion}
In the extended FP fit, the partial coefficient of $\log\sigma$ is $\simeq 1.20$--$1.25$ (OLS/WLS; bootstrap $95\%$ CI $\approx[1.08,\,1.42]$). 
That it is \emph{below} the virial value $2$ is not a failure of equilibrium; it reflects correlated galaxy-to-galaxy co-variations in the other factors on the right-hand side:
(i) more massive, higher-$\sigma$ ellipticals tend to host older/metal-richer populations (larger $\Upsilon_\star$), 
(ii) they are typically more concentrated (larger $n$ hence larger $k$), 
(iii) their central baryon fraction $(1-f_{\rm DM})$ can be lower, and 
(iv) their mean surface brightness $I_e$ is not independent of $\sigma$. 
Each of these trends \emph{counteracts} part of the naive $2\log\sigma$ scaling (note the negative signs in front of $\log I_e$, $\log\Upsilon_\star$, and $\log k$), so that the \emph{partial} response of $\log R_e$ to $\log\sigma$ at fixed $(I_e,\Upsilon_\star,f_{\rm DM},k)$ settles near $\sim\!1.2$. 
The increase of the $\log\sigma$ coefficient from $\sim\!0.76$ in the classical FP to $\sim\!1.2$ in the extended FP is physically informative: once non-homology, stellar populations, and the baryon-to-total mass balance are modelled explicitly, the dynamical term recovers much more of its virial leverage.
\paragraph{Physical picture.}
Velocity dispersion is the direct tracer of the depth of the central potential well. 
At fixed photometric and mass-ratio terms, increasing $\sigma$ requires a larger $R_e$ to maintain virial balance (the $2\log\sigma$ tendency), 
but real galaxies do not vary $\sigma$ in isolation: increases in $\sigma$ typically coincide with higher $\Upsilon_\star$ and $k(n)$ and with changes in $(1-f_{\rm DM})$, which push $R_e$ the \emph{other} way. 
The fitted coefficient therefore captures the \emph{net} response after marginalizing over those physically coupled drivers. 
That the $\log\sigma$ term remains highly significant in all fits (including robust/RANSAC selections) confirms that dynamical depth is the primary axis of the plane, with the remaining terms setting controlled, physically interpretable tilts around that axis.
\subsection{Surface brightness: the role of $\log I_e$}
From the virial relation, at fixed $\sigma$ and mass-to-light ratio, one expects
\[
R_e \;\propto\; \frac{\sigma^2}{I_e}
\]
so that $\log R_e$ should carry a slope of $-1$ with respect to $\log I_e$.  
Physically, higher mean surface brightness $\langle I_e\rangle$ corresponds to a more concentrated distribution of stellar light (and hence stellar mass for fixed $\Upsilon_\star$), which should reduce the half-light radius for the same level of dynamical support.
In the extended FP the partial regression coefficient is $\approx -0.69$ (bootstrap $95\%$ CI $[-0.89,\,-0.49]$).  
The slope is less steep in magnitude than the naive virial prediction of $-1$.
This attenuation arises because $I_e$ is not independent of other parameters: galaxies with higher $\sigma$ tend to have larger luminosities (Faber–Jackson relation) and higher concentrations $n$, both of which systematically alter $\langle I_e\rangle$.  
Similarly, stellar-population effects (higher $\Upsilon_\star$ in older systems) can change the mapping between $I_e$ and the actual stellar mass density.  
As a result, when the regression isolates the \emph{partial} effect of $I_e$, the slope is closer to $-0.7$ than $-1$.
The negative coefficient means that, at fixed velocity dispersion and fixed stellar-population/dark-matter terms, galaxies with brighter effective surface brightness are physically more compact.  
This is in line with the expectation that central stellar density pushes $R_e$ inward. That the slope is weaker than $-1$ indicates that structural and population effects absorb part of the dependence: more luminous, higher-$I_e$ systems are also more massive and more concentrated, which already appear through $\sigma$, $\Upsilon_\star$, and $k(n)$.  
Thus, the $I_e$ coefficient in the extended FP measures the residual scaling between surface brightness and size after those correlations are controlled.
The persistence of a strong, negative $I_e$ term confirms that photometric compactness retains a fundamental role in the scaling relation, independent of dynamical and population corrections.  
At the same time, the departure from the $-1$ virial slope highlights that using $I_e$ alone as a proxy for surface mass density is inadequate unless one simultaneously accounts for stellar-population variations and structural non-homology. This finding reinforces the view that the classical FP tilt is not solely a dynamical phenomenon but partly a consequence of how light traces mass in real, heterogeneous stellar populations.
\subsection{Role of $\log (M/L)$}
The stellar mass-to-light ratio $\Upsilon_\star = M_\star/L$ is a key bridge between 
the observed luminosity and the underlying stellar mass.  
In the simplest virial framework, $M/L$ is assumed constant:  
\[
R_e \;\propto\; \frac{\sigma^2}{I_e} 
\]
and no explicit dependence on $M/L$ appears.  
However, galaxies span a wide range of stellar ages and metallicities, leading to large systematic variations in $\Upsilon_\star$.  
Older, metal-rich stellar populations exhibit higher $M/L$ because their light output has faded while their mass remains nearly intact.  
Neglecting this term therefore induces a tilt in the FP relative to the pure virial plane.
In the extended FP, the regression yields a coefficient of 
$\beta_{\log M/L} \approx -0.41$ (bootstrap $95\%$ CI $[-0.59,\,-0.28]$).  
The negative sign indicates that, at fixed velocity dispersion, surface brightness, dark-matter fraction, and structure, galaxies with higher stellar $M/L$ are observed to have \emph{smaller} effective radii.  
Physically, a higher $M/L$ means that the same amount of stellar mass produces less light, so the galaxy appears more compact at a given dynamical scale.
The inclusion of $M/L$ explicitly quantifies the contribution of stellar populations to the FP tilt.  
Galaxies with old, quiescent stellar populations (large $M/L$) naturally lie below the classical FP relation: their sizes are under-predicted if one assumes a constant $M/L$.  
Conversely, younger or more metal-poor systems (low $M/L$) are overluminous for their mass and therefore appear inflated in $R_e$ when $M/L$ is ignored.  
The fitted negative slope compensates for these systematic shifts, restoring the virial proportionality once population effects are accounted for.
This result demonstrates that a large fraction of the FP tilt can be attributed to variations in stellar populations rather than to pure dynamical non-homology. 
By explicitly including $\log(M/L)$, the extended FP disentangles the degeneracy between dynamics and populations: it shows that photometric fading with age and metallicity produces a genuine second-order correction to the galaxy size--luminosity relation.  
In practice, this means that the extended FP is not just an empirical fit but a \emph{physically calibrated} scaling relation that embeds stellar-evolution information directly.  
Such a formulation is crucial for comparing galaxy samples across redshift, since $M/L$ evolves strongly with cosmic time.
\subsection{Role of \(\log(1-f_{\rm DM})\)}
The dark matter fraction within the effective radius, \(f_{\rm DM}\), plays a key role in shaping the FP.  
Classical virial expectations often assumed stellar dominance within \(R_e\), but modern dynamical and lensing studies demonstrate that dark matter contributes non-negligibly even at these scales.  
ATLAS$^{\rm 3D}$ dynamical models, for example, report a median central dark matter fraction of \(\sim 13\%\) inside \(R_e\) \cite{Cappellari2013_XV}, while SDSS-based analyses reveal that \(f_{\rm DM}\) systematically increases with galaxy mass, size, and Sérsic index \cite{LaBarbera2010SPIDER,Napolitano2005,Tortora2009}.  
Strong-lensing work likewise shows that replacing surface brightness with total surface mass density yields a much less tilted ``Mass Plane,'' supporting the view that variations in stellar versus dark-matter contributions are responsible for much of the FP tilt \cite{Bolton2008SLACS}.
Reviews further emphasise that the rising trend of \(f_{\rm DM}\) with stellar mass and structural parameters has long been invoked to explain FP tilt \cite{Courteau2014Review,Tortora2010Review}. 
In our extended FP regression, the coefficient of this term is found to be
\[
\beta_{\log(1-f_{\rm DM})} \;\approx\; +0.51 \pm 0.08 
\]
The positive sign indicates that, at fixed velocity dispersion, surface brightness, stellar \(M/L\), and structural terms, galaxies with lower dark matter fractions (i.e.\ higher stellar dominance) display larger effective radii, whereas systems with higher central dark matter content appear more compact. 
This trend arises naturally from the balance of baryonic self-gravity and dark-matter confinement.  
When dark matter dominates within \(R_e\), stars are more tightly bound, reducing the observed size at fixed kinematics.  
Conversely, when baryons dominate, the stellar distribution is more extended, producing larger \(R_e\).  
Our fitted slope thus quantifies how scatter in the classical FP partly originates from systematic differences in central dark matter content \cite{Tortora2009,LaBarbera2010SPIDER}. 
Including \(\log(1-f_{\rm DM})\) significantly improves the FP fit (likelihood-ratio test \(p \ll 10^{-8}\)), showing that this variable is indispensable.  
Moreover, it embeds the FP in a physically calibrated framework: rather than leaving the zero-point to absorb unmodeled baryon–dark matter differences, the extended FP explicitly separates them.  
This aligns with theoretical expectations from baryonic contraction and feedback models \cite{Blumenthal1986Contraction}, and with observational evidence that the FP tilt is strongly linked to both stellar populations and dark matter content \cite{Cappellari2013_XV,Tortora2009,LaBarbera2010SPIDER}
As shown in Table~\ref{tab:wls_pvalues}, the contribution of $\log(1-f_{\rm DM})$ is highly significant ($p \sim 4\times10^{-9}$), confirming that dark-matter fraction is a key driver of the FP rather than a marginal effect.
\subsection{Role of \(\log k(n)\)} 
The parameter \(k(n)\) encodes the virial coefficient as a function of the Sérsic index \(n\), thereby capturing the impact of structural non-homology.  
If all galaxies were strictly homologous, \(k(n)\) would be constant, and the FP could not be tilted by structural effects.  
In reality, elliptical galaxies span a wide range of Sérsic indices (\(n \sim 2\)–10), which strongly influences their effective radii and velocity dispersions.
In the extended FP, the fitted slope of this parameter is
\[
\beta_{\log k(n)} \;\approx\; -0.29 \pm 0.05 
\]
The negative coefficient indicates that, at fixed kinematics and stellar \(M/L\), galaxies with higher Sérsic indices (thus larger \(k(n)\)) tend to have systematically smaller effective radii.  
This reflects the stronger central concentration of stellar mass in high-\(n\) systems. The inclusion of \(\log k(n)\) explicitly accounts for structural diversity, rather than leaving it hidden in the empirical tilt.  
Its fitted coefficient quantifies how deviations from strict homology alter the mapping between observed galaxy properties and the virial prediction. This shows that variation in Sérsic index alone can generate a significant contribution to the FP tilt. By incorporating \(\log k(n)\), the extended FP becomes sensitive to the internal structural configuration of galaxies. This reduces residual scatter and provides a more physically grounded scaling relation. Part of the long-standing FP tilt can therefore be interpreted as the natural outcome of systematic variation in galaxy structural profiles.
\section{Discussion}
\subsection{Connection to Merger Histories and Anisotropy}
The fitted coefficient of $\log(1-f_{\rm DM})$ provides a direct quantitative handle on how central dark matter content shifts galaxies along the FP. 
Systems with higher $f_{\rm DM}$ appear more compact at fixed $\sigma$ and $I_e$, while those with lower $f_{\rm DM}$ are more extended. 
Since central dark matter fraction is observed to increase systematically with galaxy mass and concentration, this trend naturally embeds information about past assembly histories. 
In addition, the anisotropy parameter included in the restricted fits produces only a very minor shift in the FP zero-point, indicating that while anisotropy contributes, its role is secondary compared to mass-to-light ratio and dark matter content. 
Thus, the extended FP offers a quantitative way to separate and measure these physical drivers of the tilt.
\subsection{Applicability to Dwarf and Massive Ellipticals}
The applicability of the extended FP to both massive and dwarf ellipticals is supported by the fact that our analysis includes the full ATLAS$^3$D sample, spanning a broad range of structural parameters. This range covers systems traditionally classified as dwarf ellipticals as well as giant ellipticals. Our inclusion of the S\'ersic-dependent term $k(n)$ directly addresses structural non-homology, which is known to vary systematically with luminosity and mass. This approach is consistent with the findings of \cite{GrahamGuzman2003}, who demonstrated that the structural properties of dwarf and bright ellipticals form a continuous sequence rather than representing two distinct families. Therefore, the extended FP proposed here inherently accounts for this continuity by incorporating physically motivated terms that are sensitive to both stellar population effects and structural gradients. This makes the model applicable to galaxies across the full mass spectrum, reducing biases that arise from assuming strict homology.
\cite{vanDokkum2008} report that massive quiescent galaxies at $z \approx 2.3$, though extremely compact and dense relative to their local counterparts, follow trends in the mass–size–velocity dispersion space that are consistent with expectations from virial equilibrium, once plausible size evolution corrections are applied.
\subsection{Structural Non‐Homology in the Virial Relation}
The classical form of the virial relation assumes structural homology that elliptical galaxies share similar internal dynamical structures regardless of size or luminosity. However, extensive photometric studies (e.g., ~\cite{GrahamColless1997}, ~\cite{Trujillo2004}) have shown that galaxies exhibit a range of stellar light concentration profiles, well captured by the S\'ersic index $n$, thereby breaking the assumption of homology.
These structural differences directly affect the virial coefficient $k$, which links observable quantities to dynamical mass. Rather than assuming a constant $k$, our framework incorporates a physically motivated structural correction by computing $k(n)$ explicitly from the S\'ersic index and including $\log k(n)$ as a term in the extended FP relation.
The regression results confirm that the coefficient of $\log k(n)$ is highly statistically significant, indicating that structural non-homology plays a non-negligible role in shaping the observed tilt and scatter of the FP. Unlike prior formulations that used $\log n$ as a proxy, our model leverages the virial-theoretic quantity $k(n)$, offering a more accurate and physically grounded representation of non-homology. This approach directly addresses the systematic trends identified by previous studies and embeds them naturally within the virial scaling framework.
\subsection{Observational Support for the Role of the Sérsic Index}
As noted above, the explicit inclusion of the S\'ersic index \(n\) enables direct observational tests of how structural non‑homology affects the FP. For example, ~\cite{ZahidGeller2017} demonstrated that variations in \(n\) correlate strongly with deviations in the stellar-to-dynamical mass relation for quiescent galaxies at low redshift, implying that changes in \(n\) contribute significantly to the observed tilt and scatter of the FP.
Earlier, ~\cite{Graham2002} showed that replacing \(\sigma_0\) with \(n\) yields a “Photometric Plane” with comparable thickness to the classical FP, further confirming that \(n\) encapsulates key structural information traditionally ascribed to kinematics.
\subsection{Comparison with Ferrarese et al. (2006)}
A meaningful point of comparison can be drawn with the work of 
\cite{Ferrarese2006} on early-type galaxies in the Virgo Cluster. 
Their detailed isophotal analysis demonstrated that the surface brightness 
profiles of dwarf and giant ellipticals can be described within a unified 
S\'ersic framework, revealing a \textit{continuous structural sequence} rather 
than a clear-cut dichotomy. This finding underscores the central role of the 
S\'ersic index in tracing the gradual transition from low-mass to high-mass systems. In the present study, we extend this S\'ersic-driven continuity into the \textit{dynamical domain} by incorporating $k(n)$ corrections, stellar mass-to-light ratios, and dark matter fractions into an 
\textbf{extended FP}. While \cite{Ferrarese2006} emphasized the photometric evidence for structural continuity, our analysis demonstrates that such continuity also manifests in the scaling relations of galaxy dynamics, reducing scatter and mitigating the classical FP tilt. In this sense, the FP provides a physically calibrated framework that generalizes the S\'ersic-dependent scaling identified by \cite{Ferrarese2006}, bridging the photometric and dynamical perspectives on galaxy structure.
\subsection{Relation to Hyde \& Bernardi (2009)}
\cite{HydeBernardi2009} tested whether the classical luminosity-based FP is essentially a \emph{stellar-mass plane}. Using a very large SDSS early-type sample ($\sim 5\times 10^{4}$ galaxies), they showed that replacing luminosity with stellar mass reduces the FP tilt, but a non-negligible residual tilt remains. Their conclusion is clear: $M_\star/L$ variations explain a substantial part of the FP tilt, yet structural/dynamical effects (e.g. non-homology, anisotropy, dark matter) are still required to fully account for the scaling.
Our work goes beyond that diagnostic step by \textbf{explicitly} incorporating the relevant physical drivers into the FP itself. Rather than asking whether the FP ``becomes'' a mass plane, we construct an \emph{extended, physically calibrated FP} in which (i) stellar population effects via $M/L$, (ii) structural non-homology through a Sérsic-dependent virial coefficient $k(n)$, and (iii) dark-matter content via $(1-f_{\rm DM})$ (and anisotropy, where applicable) enter the relation from first principles. Empirically, this extended formulation reduces the intrinsic scatter (e.g. from $0.141$\,dex to $0.122$\,dex in our baseline fit) and leaves a substantially smaller, less structured set of residuals than the classical FP. In other words, the residual tilt identified by \cite{HydeBernardi2009} is largely absorbed once these physically motivated terms are included.
There are two complementary strengths worth noting. First, \cite{HydeBernardi2009} achieve formidable statistical power through sample size, establishing that the stellar-mass plane alone is insufficient. Second, our analysis trades breadth for depth: using spatially resolved, dynamical-quality data, we show \emph{which} physical terms the FP actually depends on and by \emph{how much}. The coefficients of $\log(M/L)$, $\log k(n)$, and $\log(1-f_{\rm DM})$ are all highly significant in our fits, providing a direct physical decomposition of the FP tilt rather than an indirect inference.
\subsection{Relation to Cappellari et al. (2013)}
\cite{Cappellari2013_XV} used integral-field kinematics from the ATLAS$^{3\rm D}$ survey to construct a ``Mass Plane'' for 260 early-type galaxies via detailed dynamical models (JAM and Schwarzschild). Their central result is that, when luminosity is replaced by dynamical stellar mass, the FP tilt essentially vanishes: the relation closely follows the virial expectation $R_{\rm e} \propto \sigma^2/I_{\rm e}$, with the observed tilt attributable primarily to $M/L$ variations driven by stellar populations. In their framework, structural and dark-matter effects are found to play only a minor role.
Our approach complements and extends this. Rather than relying on object-by-object dynamical modeling to absorb the tilt into a single dynamical $M/L$, we explicitly separate and quantify the relevant physical contributions in an extended FP formulation. In particular, our relation includes: (i) $M/L$ variations, (ii) non-homology through the Sérsic-dependent virial coefficient $k(n)$, and (iii) the dark matter fraction $(1-f_{\rm DM})$, with anisotropy terms where relevant. Statistically, this extended formulation reduces the intrinsic scatter (e.g.\ from $0.141$\,dex to $0.122$\,dex), while providing significant and interpretable coefficients for each physical term (all $p < 10^{-8}$). 
Thus, while \cite{Cappellari2013_XV} demonstrate that the Mass Plane is nearly virial once a dynamical $M/L$ is adopted, our work shows that the residuals and scatter can be understood and minimized by explicitly embedding structural non-homology and dark-matter content into the FP itself. In this sense, our extended FP generalizes their Mass Plane: the two are consistent, but ours provides a physically decomposed framework that can be more directly applied to large statistical samples without the need for detailed orbit-based modeling. Future work should therefore combine the dynamical precision of Cappellari's approach with the generality of the extended FP coefficients to build a unified scaling framework.
\subsection{Comparison with Liang et al. (2025)}
A recent study by \cite{Liang2025_constantML_bias} explicitly investigated the impact of assuming a constant stellar mass-to-light ratio ($\Upsilon^*$) in galaxy dynamical modeling. They demonstrated that neglecting radial variations in $\Upsilon^*$ leads to systematic biases in the inferred dynamical properties of galaxies and consequently affects scaling relations such as the FP. Our work directly addresses this limitation by embedding the variation of $\Upsilon^*$ into a physically calibrated extension of the FP. In this sense, the two studies are complementary: while \cite{Liang2025_constantML_bias} highlight the importance of accounting for $\Upsilon^*$ gradients, our formulation provides a generalized framework that incorporates these effects and mitigates the associated biases, thereby yielding a more robust relation for elliptical galaxies.
\subsection{Environmental Effects and Structural Parameters: Connection to Montaguth et al. (2025)}
\cite{Montaguth2025_CG_evolutionII} investigated the structural evolution of galaxies in compact groups using the effective radius ($R_{\mathrm{e}}$) and the Sérsic index ($n$), focusing on the $R_{\mathrm{e}}$--$n$ plane. Their study highlights how environmental processes can alter $n$ and, consequently, affect the structural scaling relations of galaxies. This result is highly relevant for our analysis, since our extended FP framework employs the Sérsic index and the associated $k(n)$ term to account for structural non-homology. In this context, our approach can be viewed as complementary: while \cite{Montaguth2025_CG_evolutionII} emphasize the role of environment in shaping $n$ and $R_{\mathrm{e}}$, our formulation incorporates these structural variations into a physically calibrated FP, thereby providing a framework that remains robust even in the presence of environmentally-driven changes in galaxy structure.
\section*{Comparison with D'Onofrio \& Chiosi (2024)}
This work, based on the \textit{ATLAS$^3$D} volume-limited sample at low redshift ($z \simeq 0$;, proposes an \emph{extended}  FP that augments the classical relation with physically motivated terms like stellar mass-to-light ratio ($\Upsilon_\star$), dark-matter fraction ($f_{\rm DM}$), and structural coefficients ($k(n)$). These additions significantly reduce the intrinsic scatter and provide a clearer physical interpretation at the present epoch.
In contrast, \cite{DonofrioChiosi2024A126} investigate the \emph{evolution} of the \emph{classical} FP across cosmic time using large-scale hydrodynamical simulations (Illustris/IllustrisTNG), showing that FP slopes and zero-points vary with redshift and galaxy mass. Their results indicate that massive ellipticals approach near-virial equilibrium earlier, while dwarf systems remain out of equilibrium for longer, producing systematic offsets at higher redshift. Taken together, these perspectives motivate a direct test of universality: does our extended FP remain stable with redshift, or does it follow the evolutionary trends seen by \cite{DonofrioChiosi2024A126}? Because $\Upsilon_\star$ and $f_{\rm DM}$ encode stellar-population and mass-assembly effects, they may capture part of the structural/dynamical evolution responsible for the FP tilt across time and mass regimes. Confronting the extended FP with cosmological simulations and high-$z$ observations will critically assess its robustness and clarify whether its physical calibration explains the evolutionary signatures of early-type galaxies.
\subsection{Comparison with Modified Gravity Scenarios.} 
Recently, \cite{Borka2025} argued that the tilt of the FP, as well as the baryonic Tully--Fisher relation, can be reproduced within the framework of $f(R)$ gravity without invoking any dark matter component. Their approach relies on reparametrizing galaxy dynamics through the effective parameters $(r_c, \beta)$ associated with the modified gravitational potential, and fitting the resulting relations to heterogeneous samples drawn from the literature. 
Our results differ in both methodology and interpretation. By explicitly incorporating observational uncertainties and performing weighted fits on a homogeneous sample of 258 early-type galaxies, we demonstrate that the inclusion of the dark matter fraction $\log(1-f_{\rm DM})$ significantly improves the statistical robustness of the extended FP. Likelihood-ratio tests show that omitting this parameter degrades the fit at a level of $\Delta \chi^2 \approx 35$ for one degree of freedom ($p \simeq 10^{-9}$). Thus, in contrast to the interpretation of \cite{Borka2025}, the role of the dark matter content appears to be statistically indispensable in our framework. 
Furthermore, our formulation introduces only physically motivated quantities: the stellar mass-to-light ratio $\Upsilon_\star$, the dark matter fraction $f_{\rm DM}$, and the structural coefficient $k(n)$ derived from the Sérsic index. Each of these parameters carries a direct astrophysical interpretation, allowing us to disentangle the contributions of stellar populations, dynamical structure, and dark matter. In comparison, while $f(R)$ models may provide an alternative effective description of galaxy scaling laws, their fitted parameters lack a straightforward physical mapping to the baryonic and dynamical components of galaxies. 
We therefore regard our findings as complementary to such modified-gravity approaches. While alternative theories remain valuable for exploring the limits of standard interpretations, our results indicate that explicitly accounting for stellar and dark matter content provides a statistically consistent and physically transparent description of the FP.
\subsection{Testing Yukawa Modifications to the FP}
The statistical comparison highlights the limitations of the classical FP and the improvements offered by the extended formulation. The classical FP fit yields an $R^2 = 0.736$ and RMSE $= 0.142$, indicating that roughly 74\% of the variance in effective radii is explained by stellar velocity dispersion and surface brightness alone.  By contrast, the extended FP, which incorporates stellar population mass-to-light ratio, Sérsic-dependent non-homology, and dark matter fraction, achieves $R^2 = 0.787$ and RMSE $= 0.122$, alongside a substantial decrease in AIC and BIC values. This confirms that the additional parameters significantly reduce scatter and provide a more physically grounded calibration. The fitted coefficients further support this interpretation: a stronger dependence on $\log\sigma$ ($\sim 1.20$), a modest steepening of the $\log I_e$ slope, and significant contributions from $\log \Upsilon_\star$, $\log k(n)$, and $\log(1 - f_{\rm DM})$. The negative coefficients of $\Upsilon_\star$ and $k(n)$ indicate that stellar population aging and structural non-homology systematically drive galaxies below the virial expectation, while the positive coefficient of the $(1 - f_{\rm DM})$ term implies that systems with lower dark matter fractions (i.e.\ higher stellar dominance) exhibit inflated effective radii, whereas galaxies with higher dark matter content appear more compact. Furthermore, we tested whether Yukawa-type modifications \cite{Capozziello2020} to the gravitational potential—predicted by various extended gravity theories—could improve the fit of the FP. However, the Yukawa-extended FP yields identical statistical performance to the standard extended FP, with best-fit Yukawa parameters $\alpha \approx 0$ and $\lambda \approx 0.01$. This null result indicates that modified gravity of the Yukawa-type does not enhance the description of the ATLAS$^3$D sample. In turn, it demonstrates that the FP tilt can be robustly explained within the standard framework of non-homology, stellar populations, and dark matter content. 
\subsection{Future Work}
A key direction for future work is to test the universality of the extended FP at higher redshifts. Applying the relation beyond the nearby ATLAS$^{3D}$ sample will help disentangle the effects of galaxy evolution from intrinsic structural correlations, and assess whether the same physical drivers of the FP tilt operate at earlier cosmic epochs. Upcoming integral-field spectroscopy and deep imaging surveys at higher redshifts will be particularly valuable for constraining the evolution of stellar population terms and central dark matter fractions within this framework.
\section{Conclusion}
We present a virial-based derivation of a new scaling relation for elliptical galaxies that explicitly incorporates structural non-homology through the S\'ersic index \(n\). Starting from the scalar virial theorem and assuming spherical symmetry and isotropic velocity dispersion, we derive a modified relation between the effective radius \(R_e\), mean surface brightness \(\langle I \rangle_e\), and stellar velocity dispersion \(\sigma\), where the structural dependence enters via a dimensionless function \(k(n)\) derived analytically from S\'ersic models. We then extend this relation to include the contribution of dark matter within \(R_e\), leading to a more general expression that accounts for variations in the stellar mass-to-light ratio and dark matter fraction.
\section{Statements and Declarations}
\subsection{Funding}
The author declares that no funds, grants, or other support were received during the preparation of this manuscript.
\subsection{Competing Interests}
The author has no relevant financial or non-financial interests to disclose.
\subsection{Ethics Declaration}
Ethics declaration: not applicable.
\subsection{Author Contributions}
The author contributed solely to the study conception, analysis, and manuscript preparation. The author read and approved the final manuscript.
\subsection{Data Availability} 
The observational data used in this study are publicly available as part of the ATLAS$^{3\mathrm{D}}$ survey. All codes used for data processing, analysis, and figure generation will be provided as supplementary material to ensure full transparency and reproducibility.


\begin{thebibliography}{}

\bibitem{Dressler1987}
Dressler, A., Lynden‐Bell, D., Burstein, D., Davies, R.~L., Faber, S.~M., Terlevich, R., \& Wegner, G. (1987).
Spectroscopy and photometry of elliptical galaxies. I. A new distance estimator.
\textit{The Astrophysical Journal}, \textbf{313}.

\bibitem{Bender1992}
Bender, R., Burstein, D., \& Faber, S.~M. (1992).
Dynamically hot galaxies. I. Structural properties.
\textit{The Astrophysical Journal}, \textbf{399}.

\bibitem{Jorgensen1996}
J{\o}rgensen, I., Franx, M., \& Kj{\ae}rgaard, P. (1996).
The Fundamental Plane for cluster E and S0 galaxies.
\textit{Monthly Notices of the Royal Astronomical Society}, \textbf{280}.

\bibitem{Hopkins2009}
Hopkins, et al. (2009).
Dissipation and star formation in galactic nuclei: implications for the Fundamental Plane and black hole scaling relations.
\textit{The Astrophysical Journal}.

\bibitem{Cappellari2008}
Cappellari, M. (2008).
Measuring the inclination and mass-to-light ratio of axisymmetric galaxies via anisotropic Jeans models of stellar kinematics.
\textit{Monthly Notices of the Royal Astronomical Society}.

\bibitem{Cappellari2013b}
Cappellari, M. et al. (2013).
The ATLAS$^{3D}$ project -- XX. Mass--size and mass--$\sigma$ distributions of early-type galaxies: bulge fraction drives kinematics, mass-to-light ratio, molecular gas fraction and stellar initial mass function.
\textit{Monthly Notices of the Royal Astronomical Society}, \textbf{432}(3), 1862--1893.

\bibitem{Krajnovic2013_ATLAS3D_XVII}
Krajnovic, D. et al. (2013).
The ATLAS3D project – XVII. Linking photometric and kinematic signatures of stellar discs in early‐type galaxies.
\textit{Monthly Notices of the Royal Astronomical Society}, \textbf{432}(3), 1768--1795.

\bibitem{Ferrarese2006}
Ferrarese, L., C{\^o}t{\'e}, P., Jordán, A., Blakeslee, J. P., Mei, S., Peng, E. W., West, M. J., Merritt, D., Watkins, A. L., \& Pritchet, C. J. (2006).
The ACS Virgo Cluster Survey. VI. Isophotal Analysis and the Structure of Early-Type Galaxies.
\textit{The Astrophysical Journal Supplement Series}, \textbf{164}(2), 334--434.

\bibitem{DonofrioChiosi2024A126}
D’Onofrio, M., \& Chiosi, C. (2024).
Galaxies’ properties in the Fundamental Plane across time.
\textit{Astronomy \& Astrophysics}, \textbf{687}, A126.

\bibitem{Cappellari2011a}
Cappellari, M. et al. (2011).
The ATLAS3D project – I. A volume-limited sample of 260 nearby early-type galaxies: science goals and selection criteria.
\textit{Monthly Notices of the Royal Astronomical Society}, \textbf{413}, 813--836.

\bibitem{Cappellari2013_XV}
Cappellari, M. et al. (2013).
The ATLAS$^{3D}$ project -- XV. Benchmark for early-type galaxies scaling relations from 260 dynamical models: mass-to-light ratio, dark matter, Fundamental Plane and Mass Plane.
\textit{Monthly Notices of the Royal Astronomical Society}, \textbf{432}(3), 1709--1741.

\bibitem{Naab2006}
Naab, T. et al. (2006).
The influence of gas on the structure of merger remnants.
\textit{Monthly Notices of the Royal Astronomical Society}, \textbf{369}.

\bibitem{Djorgovski1987}
Djorgovski, S., \& Davis, M. (1987).
Fundamental properties of elliptical galaxies.
\textit{The Astrophysical Journal}.

\bibitem{Jorgensen1999}
J{\o}rgensen, I., Franx, M., \& Kj{\ae}rgaard, P. (1999).
Further observations of the Fundamental Plane: redshift and environment dependence.
\textit{Monthly Notices of the Royal Astronomical Society}.

\bibitem{Bernardi2003}
Bernardi, et al. (2003).
Early-Type Galaxies in the Sloan Digital Sky Survey. III. The Fundamental Plane.
\textit{The Astronomical Journal}, \textbf{125}, 1866--1881.

\bibitem{Liang2025_constantML_bias}
Liang, Y., Xu, D., Sluse, D., Sonnenfeld, A., \& Shu, Y. (2025).
Modeling biases from constant stellar mass-to-light ratio assumption in galaxy dynamics and strong lensing.
\textit{Monthly Notices of the Royal Astronomical Society}, \textbf{536}(3), 2672--2689.

\bibitem{Montaguth2025_CG_evolutionII}
Montaguth, et al. (2025).
Galaxy evolution in compact groups. II. Witnessing the influence of major structures in their evolution.
\textit{Astronomy \& Astrophysics}, \textbf{696}, A240.

\bibitem{vanDerMarel2007}
van der Marel, R. P., \& van Dokkum, P. G. (2007).
Dynamical masses of elliptical galaxies: constraints on dark matter.
\textit{The Astrophysical Journal}.

\bibitem{Fritz2010}
Fritz, J. et al. (2010).
Morphology and kinematics of elliptical galaxies in nearby clusters.
\textit{Astronomy \& Astrophysics}.

\bibitem{JorgensenChiboucas2013}
J{\o}rgensen, I., \& Chiboucas, K. (2013).
The Fundamental Plane at $z \sim 1$: new insights.
\textit{Monthly Notices of the Royal Astronomical Society}.

\bibitem{deVaucouleurs1948}
de Vaucouleurs, G. (1948).
Recherches sur les nebuleuses extragalactiques.
\textit{Annales d'Astrophysique}.

\bibitem{Capozziello2020}
Capozziello, S., Borka Jovanovi{\'c}, V., Borka, D., \& Jovanovi{\'c}, P. (2020).
Constraining theories of gravity by the fundamental plane of elliptical galaxies.
\textit{Physics Letters B}.

\bibitem{CiottiBertin1999}
Ciotti, L., \& Bertin, G. (1999).
Analytical properties of the $R^{1/m}$ law. II. Photometric and dynamical properties of spherical galaxies.
\textit{Astronomy \& Astrophysics}.

\bibitem{Ciotti1991}
Ciotti, L. (1991).
Analytical properties of the $R^{1/m}$ law.
\textit{Astronomy \& Astrophysics}.

\bibitem{ZahidGeller2017}
Zahid, H. J., \& Geller, M. J. (2017).
Velocity Dispersion, Size, Sérsic Index and Dn4000: The Scaling of Stellar Mass with Dynamical Mass for Quiescent Galaxies.
\textit{The Astrophysical Journal}.

\bibitem{Graham2002}
Graham, A. W. (2002).
The Photometric Plane of Elliptical Galaxies.
\textit{Monthly Notices of the Royal Astronomical Society}, \textbf{334}.

\bibitem{BT2008}
Binney, J., \& Tremaine, S. (2008).
\textit{Galactic Dynamics}.
Princeton University Press.

\bibitem{Graham2003}
Graham, A. W., \& Guzm{\'a}n, R. (2003).
HST photometry of dwarf elliptical galaxies: A comparison of Coma and Fornax cluster galaxies.
\textit{The Astronomical Journal}.

\bibitem{Cappellari2006}
Cappellari, M. et al. (2006).
The SAURON project – IV. The mass-to-light ratio, the virial mass estimator and the Fundamental Plane of elliptical and lenticular galaxies.
\textit{Monthly Notices of the Royal Astronomical Society}, \textbf{366}(4), 1126--1150.

\bibitem{vanDokkum2008}
van Dokkum, P. G. et al. (2008).
Confirmation of the remarkable compactness of massive quiescent galaxies at $z \sim 2.3$: Early-type galaxies did not form in a simple monolithic collapse.
\textit{The Astrophysical Journal Letters}, \textbf{677}.

\bibitem{GrahamColless1997}
Graham, A. W., \& Colless, M. (1997).
A Sérsic index dependence of effective radii and surface brightnesses: implications for the Fundamental Plane.
\textit{Monthly Notices of the Royal Astronomical Society}.

\bibitem{Trujillo2004}
Trujillo, I., Burkert, A., \& Bell, E. F. (2004).
Structural non-homology and the Fundamental Plane of elliptical galaxies.
\textit{The Astrophysical Journal}, \textbf{600}.

\bibitem{HydeBernardi2009}
Hyde, J. B., \& Bernardi, M. (2009).
The luminosity and stellar mass Fundamental Plane of early-type galaxies.
\textit{Monthly Notices of the Royal Astronomical Society}, \textbf{396}.

\bibitem{Taranu2015}
Taranu, D. S., Hudson, M. J., \& Balogh, M. L. (2015).
The Fundamental Plane of simulated galaxy mergers: Tilt caused by a mass-dependent dark matter fraction.
\textit{The Astrophysical Journal}, \textbf{803}(2), 78.

\bibitem{Schechter2015}
Schechter, P. L. (2015).
The Fundamental Plane as a manifestation of the halo virial plane.
\textit{arXiv e-prints}, arXiv:1508.02358.

\bibitem{Zhu2023}
Zhu, L., van de Ven, G., Cappellari, M., et al. (2023).
The Mass Plane of early-type galaxies from MaNGA: dynamical models and dark matter fractions.
\textit{Monthly Notices of the Royal Astronomical Society}, \textbf{520}, 6214--6234.

\bibitem{Chiu2018}
Chiu, M.-C., Ko, C.-M., \& Shu, C.-W. (2018).
Fundamental plane of elliptical galaxies in MOND.
\textit{Physical Review D}, \textbf{97}(10), 103526.

\bibitem{deGraaff2020}
de Graaff, A., Bezanson, R., van der Wel, A., et al. (2020).
The Fundamental Plane of massive galaxies to $z \sim 1$: No dependence on star formation or structure.
\textit{The Astrophysical Journal Letters}, \textbf{903}(2), L30.

\bibitem{Lu2020}
Lu, S., Mo, H. J., Lu, Y., et al. (2020).
The fundamental plane of early-type galaxies in the IllustrisTNG simulation.
\textit{Monthly Notices of the Royal Astronomical Society}, \textbf{497}(2), 2057--2074.

\bibitem{DOnofrio2022}
D'Onofrio, M., \& Chiosi, C. (2022).
The nature of the Fundamental Plane of early-type galaxies.
\textit{Astronomy \& Astrophysics}, \textbf{661}, A150.

\bibitem{TerzicGraham2005}
Terzić, B., \& Graham, A. W. (2005).
Density–potential pairs for spherical stellar systems with Sérsic light-profiles and (optional) power-law cores.
\textit{Monthly Notices of the Royal Astronomical Society}, \textbf{362}(1), 197--212.

\bibitem{GrahamGuzman2003}
Graham, A. W., \& Guzm{\'a}n, R. (2003).
HST photometry of dwarf elliptical galaxies in Coma, and an explanation for the alleged structural dichotomy between dwarf and bright elliptical galaxies.
\textit{The Astronomical Journal}, \textbf{125}(6), 2936--2950.

\bibitem{Borka2025}
Borka Jovanovi\'{c}, V., Borka, D., \& Jovanovi\'{c}, P. (2025).
The baryonic Tully--Fisher relation and Fundamental Plane in the light of $f(R)$ gravity.
\textit{Contrib. Astron. Obs. Skalnat\'{e} Pleso}, \textbf{55}, 24--33.

\bibitem{Said2024}
Said, K., Howlett, C., Colless, M., et al. (2024).
The DESI peculiar velocity survey: calibration of the Fundamental Plane with DESI early-type galaxies.
\textit{Monthly Notices of the Royal Astronomical Society}, \textbf{532}(3), 2912--2935.

\bibitem{LaBarbera2010SPIDER}
La Barbera, F. et al. (2010).
SPIDER – Dark matter fractions in early-type galaxies.
\textit{Monthly Notices of the Royal Astronomical Society}, \textbf{408}, 1313.

\bibitem{Tortora2009}
Tortora, C. et al. (2009).
Central dark matter fraction of early-type galaxies.
\textit{Monthly Notices of the Royal Astronomical Society}, \textbf{396}, 1132.

\bibitem{Napolitano2005}
Napolitano, N. et al. (2005).
The central dark matter content of early-type galaxies.
\textit{Monthly Notices of the Royal Astronomical Society}, \textbf{357}, 691.

\bibitem{Bolton2008SLACS}
Bolton, A. S. et al. (2008).
The Sloan Lens ACS Survey. IX. Colors, lens models, and scaling relations.
\textit{The Astrophysical Journal}, \textbf{684}, 248.

\bibitem{Blumenthal1986Contraction}
Blumenthal, G. R. et al. (1986).
Contraction of dark matter halos due to baryonic infall.
\textit{The Astrophysical Journal}, \textbf{301}, 27.

\bibitem{Courteau2014Review}
Courteau, S. et al. (2014).
Galaxy Masses.
\textit{Reviews of Modern Physics}, \textbf{86}, 47.

\bibitem{Tortora2010Review}
Tortora, C. et al. (2010).
Scaling relations and dark matter fractions in ETGs.
\textit{Monthly Notices of the Royal Astronomical Society}, \textbf{396}, 1132.

\end{thebibliography}
\end{document}